%% file: power_studies_multivariate.tex
\documentclass[]{article}
\usepackage{amsmath,amssymb}
\usepackage{lmodern}
\usepackage[table]{xcolor}
\usepackage{colortbl}
\usepackage[margin=1in]{geometry}
\usepackage{color}
\usepackage{fancyvrb}
\usepackage{graphicx}
\usepackage[hypertexnames=false]{hyperref}
\usepackage{booktabs}
\usepackage{float}
\title{Power Studies For Two-Sample and \newline Goodness-of-Fit  Methods For Multivariate Data}
\author{Wolfgang Rolke}
\date{
  University of Puerto Rico - Mayaguez
	\newline
	\texttt{wolfgang.rolke@upr.edu} 
  \newline	
  \today	
}

\begin{document}
\maketitle

\begin{abstract}
   We present the results of a large number of simulation studies regarding the power of various goodness-of-fit as well as nonparametric two-sample tests for multivariate data.  In two dimensions this includes both continuous and discrete data, in higher dimensions continuous data only. In general no single method can be relied upon to provide good power, any one method may be quite good for some combination of null hypothesis and alternative and may fail badly for another. Based on the results of these studies we propose a fairly small number of methods chosen such that for any of the case studies included here at least one of the methods has good power. 
	The studies were carried out using the R packages  \emph{MD2sample} and \emph{MDgof}, available from CRAN. 
\end{abstract}

\section{Introduction}\label{introduction}

Both the goodness-of-fit (gof) and the nonparametric two-sample problem
have histories going back a century, with many contributions by some of
the most eminent statisticians. In the goodness-of-fit problem we have a
sample \((\mathbf{x}_1,..,\mathbf{x}_n)\) drawn from some a random vector $\mathbf{X}$ with probability distribution \(F\),
possibly with unknown parameters, and we wish to test \(H_0:\mathbf{X}\sim F\).
In the two-sample problem we also have a second sample \((\mathbf{y}_1,..,\mathbf{y}_m)\)
from some distribution \(G\), and here we want to test \(H_0:F=G\), that
is we want to test whether the two data sets were generated by the same
(unspecified) distribution.

The literature on both of these problems is vast and steadily growing.
Detailed discussions can be found in \cite{agostini1986},
\cite{thas2010}, \cite{raynor2009}. For an introduction to
Statistics and hypothesis testing in general see \cite{casella2002} or \cite{bickel2015}.

The power studies in this article were carried out using \textbf{R}
programs in the packages \emph{MD2sample} and \emph{MDgof}, available from
the CRAN website. In the case of univariate data many methods such as the Kolmogorov-Smirnov test are
already implemented for both problems in base \textbf{R}.  For multivariate data however the 
available software is much smaller, and most of the methods included 
in \emph{MD2sample} and \emph{MDgof} are not implemented anywhere else.

The highlights of both those packages are:

\begin{itemize}
\item
  many methods are implemented for continuous data in any dimension and discrete
  data in two dimensions. Discrete data includes the case of histogram (aka discretized or binned) data.\\
\item
  the methods are implemented using both \emph{Rcpp} \cite{rcpp2024} and
  parallel programming.\\
\item
  the routines allow for a random sample size, assumed to come from a
  Poisson distribution.\\
\item
  there are routines that allow the user to combine several tests and
  find a corrected p value.\\
\item
  the routines can also use any other user-defined tests.\\
\item
  the packages include routines to easily carry out power studies.
\item
  they include routines to easily run a large number of power studies and thereby compare a new method's performance to those included in the packages.	
\end{itemize}

Including tests for discrete data in two dimensions is useful in two ways: Of course
discrete data is of interest in its own right, and there are no
implementations in \emph{R} at this time. It also makes it possible to apply
the tests to very large continuous data sets via discretization. While a
test for a continuous data set with (say) 10000 observations each can
be done in a matter of a few minutes, for larger data sets the
calculations will be quite time consuming. Binning the data and then
running the corresponding discrete tests however is quite fast. At least this
is true for bivariate data. In more than two dimensions the number of bins
grows rapidly, and the number of observations required would be extremely large. We 
therefor implemented discrete methods in two dimensions only.

There are also situations where the underlying distribution is
continuous but the data is collected in binned form. This is for example
often the case for data from high energy physics experiments and from
astronomy because of finite detector resolution. In this situation the
theoretical distribution is continuous but the data is discrete.

For the tests in the two-sample problem p-values are found via the
permutation method.  In the
goodness-of-fit case p-values are always found via simulation (aka parametric bootstrap). 

\section{The Types of Problems Included in this
Study}\label{the-types-of-problems-included-in-this-study}

In general goodness-of-fit and two sample methods have found many uses. For example, in high energy physics they are sometimes used for discoveries. Our aim here is to use them for their original purpose. That is, a researcher has a data set and wishes to do some inference. The method he wishes to employ requires a probability model, and he has one he believes will fit. But before continuing his analysis he wishes to make certain that his model, in the immortal words of George Box, is ``good enough''.

The kinds of situations in which these methods will be useful are:

\begin{itemize}
\item
  \textbf{Goodness-of-Fit Problem - Continuous Data}: We have a sample
  \(x\) of size of \(n\). \(F\) is a continuous probability
  distribution, which may depend on unknown parameters. We want to test
  \(X\sim F\).
\item
  \textbf{Goodness-of-Fit Problem - Discrete Data}: We have a set of
  values \emph{valsx} of one variable and \emph{valsy} of the other variable, and a vector of counts \(x\). \(F\) is a discrete
  probability distribution, which may depend on unknown parameters. We
  want to test \(X\sim F\).
\item
  \textbf{Two-sample Problem - Continuous Data}: We have a sample \(x\)
  of size of \(n\), drawn from some unknown continuous probability
  distribution \(F\), and a sample \(y\) of size \(m\), drawn from some
  unknown continuous probability distribution \(G\). We want to test
  \(F=G\).
\item
  \textbf{Two-sample Problem - Discrete Data}: We have a set of
  values \emph{valsx} of one variable and \emph{valsy} of the other variable and vectors of counts \(x\) and \(y\), drawn from some
  unknown discrete probability distributions \(F\) and \(G\). We want to
  test \(F=G\).
\end{itemize}

For a goodness-of-fit test we also have the case of a simple test, where 
the distribution function is fully specified (say $\mathbf{X}\sim \mathbf{N}(\mathbf{\mu}, \mathbf{\Sigma})$) 
or the case of a composite test where some parameters are unspecified and have to be estimated
from the data.

\section{Goodness-of-fit/two-sample hybrid tests}

There is another (quite popular) way to carry out a goodness-of-fit test. In these tests one generates a second data set $y$ under the null hypothesis, using a parametric bootstrap approach if parameters are estimated. Then one carries out a two-sample test.

There are a number of reasons one might do this:
\begin{itemize}
\item 
One can now employ any two-sample method.   
\item
The actual distribution function $F$ might not be calculable, maybe because it requires high-dimensional integration.  
\end{itemize}

This approach does raise an additional issue. If generating data from $F$ is fast, how large a sample should be generated, relative to the sample size of the real data? Generally one would expect the power to increase with a larger data set. 

In this paper we will include this hybrid method with two sample sizes for the MC data set, one equal to the size of the real data and another five times as large.

\section{Highlights of the Results}

Detailed results of our studies can be found later, here are some highlights:

\begin{itemize}
\item
  While there are a plethora of methods for the two-sample problem, multivariate goodness-of-fit tests are much more rare. In fact we found only two in the literature, namely tests by Bickel and Breiman and by Bakshaev and Rudskis. Some more methods are also included in \emph{MDgof} but work only in two dimensions.
\item
  As one would expect, no single method is uniformly better than all the others. 
\item
  If the null hypothesis is true at least for the marginal distributions, that is, if those are indeed the same under the null hypothesis and the alternative, then in both the goodness-of-fit and the two-sample problem the classic chi-square test (with just a few bins!) has quite impressive power, often much better than any other method. Unfortunately chi-square tests require binning and are really only feasible in two dimensions.
\item
  Hybrid methods (doing a goodness-of-fit test by generating a second MC data set) require	a larger MC data set than the real data in order for any method to be competitive, and then not by much. So as long as it is possible to do a classic goodness-of-fit test, these are preferred.	
\item
  In a hybrid test the power can usually be improved by using a larger MC data set, if this is feasible. Just how much larger is a difficult question to answer.	
\item
  For continuous data and a two-sample problem (or a hybrid problem if it is not possible to do a goodness-of-fit test because calculating values from the cumulative distribution function is not possible) the MMD test is the single best 	option. For a goodness-of-fit problem in 2 dimensions a chi-square test with a small number of bins works well, especially in cases where the distributions of the marginals are indeed those from the null hypothesis. In higher dimensions several tests have about equal power. 
\end{itemize} 

Based on the results of these simulation studies our recommendation for the best methods are:

\begin{itemize}
\item
  Goodness-of-fit, continuous data, dim=2: Bakshaev-Rudzkis, Fasano-Fraceschini, Ripley's K, a chi-square test with equal probability bins and a small grid (say 5 by 5), and the simplified versions of Kuiper's test, Anderson-Darling test and Cramer-vonMises test implemented in \emph{Mdof}, see the section on Methods.
\item		
  Goodness-of-fit, continuous data, dim $>2$:  Bakshaev-Rudzkis and the simplified versions of Kuiper's test, Anderson-Darling test and Cramer-vonMises. Note that these are the same as for the dim=2 case except that some of those methods are now no longer available.
\item 	
  Goodness-of-fit, discrete data: Anderson-Darling, Kuiper, Kullback-Leibler and classic Pearson chi-square.
\item
  Goodness-of-fit + hybrid methods, dim=2: EP chisquare,  Fasano-Fraceschini, MMD with MC data set 5 times the size of real data set.
\item
  Goodness-of-fit + hybrid methods, dim>2: qCvM as well as Biswas-Ghosh, 5-nearest-neighbor and MMD, all with MC data set 5 times the size of real data set.	
\item
  Two-sample, continuous data, dim=2: Aslan-Zech, Biswas-Ghosh, MMD.	
\item 
  Two-sample, continuous data, dim>2: Biswas-Ghosh, MMD, 5-nearest-neighbor.	
\item
  Two-sample, discrete data, dim=2: Aslan-Zech, Anderson-Darling, Kuiper and a chi-square test.	
\end{itemize}

\section{The Methods}\label{the-methods}

\subsection{Chi-square tests}

\subsubsection{Continuous data - goodness-of-fit}

Because of the need for binning chi-square tests are implemented for bivariate
data only. The data is binned into a rectangular grid, with $n_1$ bins in the 
$x$ direction and $n_2$ bins into the y direction. The simulations all use
$n_1=n_2=5$. The bins are either equal-space \emph{ES} or equal probability \emph{EP}.

In all cases neighboring bins with low counts are joined until all bins
have an expected count of at least 5 in the goodness-of-fit problems and an actual count of 5 in the two-sample problem. In all cases the p-values are found using
the usual chi-square approximation, except in the goodness-of-fit discrete data case, where simulation is used.

If parameters have to be estimated, this is done via the user-provided
routine \(phat\). As long as the method of estimation used is consistent
and efficient and the expected counts are large enough the chi-square
statistic will have a chi-square distribution, as shown by (Fisher 1922)
and (Fisher 1924).

The formula for the test statistic is

$$\sum_{i,j} \frac{(O_{ij}-E_{ij})^2}{E_{ij}}$$

where $O_{ij}$ is the observed count in bin $(i,j)$ and $E_{ij}$ is the expected count under the null hypothesis.

\subsubsection{Discrete data - goodness-of-fit}

Here the binning is already given. In this case the p-value is found via simulation rather than the usual large sample theory because there are several tests using the same ingredients ($O_{ij}$ and $E_{ij}$) but without a large sample theory, so using simulation is needed anyway. This also means that joining bins so that $E_{ij}>5$ (say) is not necessary.

\subsubsection{Continuous data - two-sample problem}

Again using the binning scheme discussed above we now have the observed counts $O_{ij}$ and $M_{ij}$ for the two data sets in the $(i,j)^{th}$ bin. Let $N_1=\sum O_{ij}$, $N_2=\sum M_{ij}$ and $N=N_1+N_2$ as well as $Z_{ij}=O_{ij}+M_{ij}$, then the chi-square test statistic is given by

$$\sum_{i,j} \frac{(O_{ij}-NZ_{ij}/N_1)^2}{NZ_{ij}/N_1}+\sum_{i=1}^{k} \frac{(M_{ij}-(NZ_{ij}/N_2)^2}{NZ_{ij}/N_2}$$

Here the "expected counts" are found by simply combining the two data sets.

Under the null hypothesis this test statistic will have a chi-square distribution with the degrees of freedom equal to the number of bins - 1.

\subsubsection{Discrete data - two-sample problem}

Similar to the continuous case with the binning already given.

\subsubsection{Binning}

There are two basic binning schemes, equal-size and equal-probability bins. Which is better depends on the case study. In many of those included here under the null hypothesis the marginals have a uniform distribution, and so these types of bins are the same.

As far the as the number of bins is concerned, previous studies have shown that fairly few bins usually lead to better power, see for example \cite{rolke2021}. Here is the result of a study that shows that the same is true for multivariate data. We use the case studies for continuous data in D=2 without estimation. We run the chisquare test with $i\times i=i^2$ bins, $i=2,..,20$. Figure 6 shows the powers, together with smoothed regression curves, which clearly shows that in almost all cases fewer bins are better. Note that the degrees of freedom is sometimes a bit smaller as in some case studies there are empty bins by design. In both \emph{MDgof} and \emph{MD2sample} the default is a $5 \times 5$ grid, but the user can change this.

\includegraphics{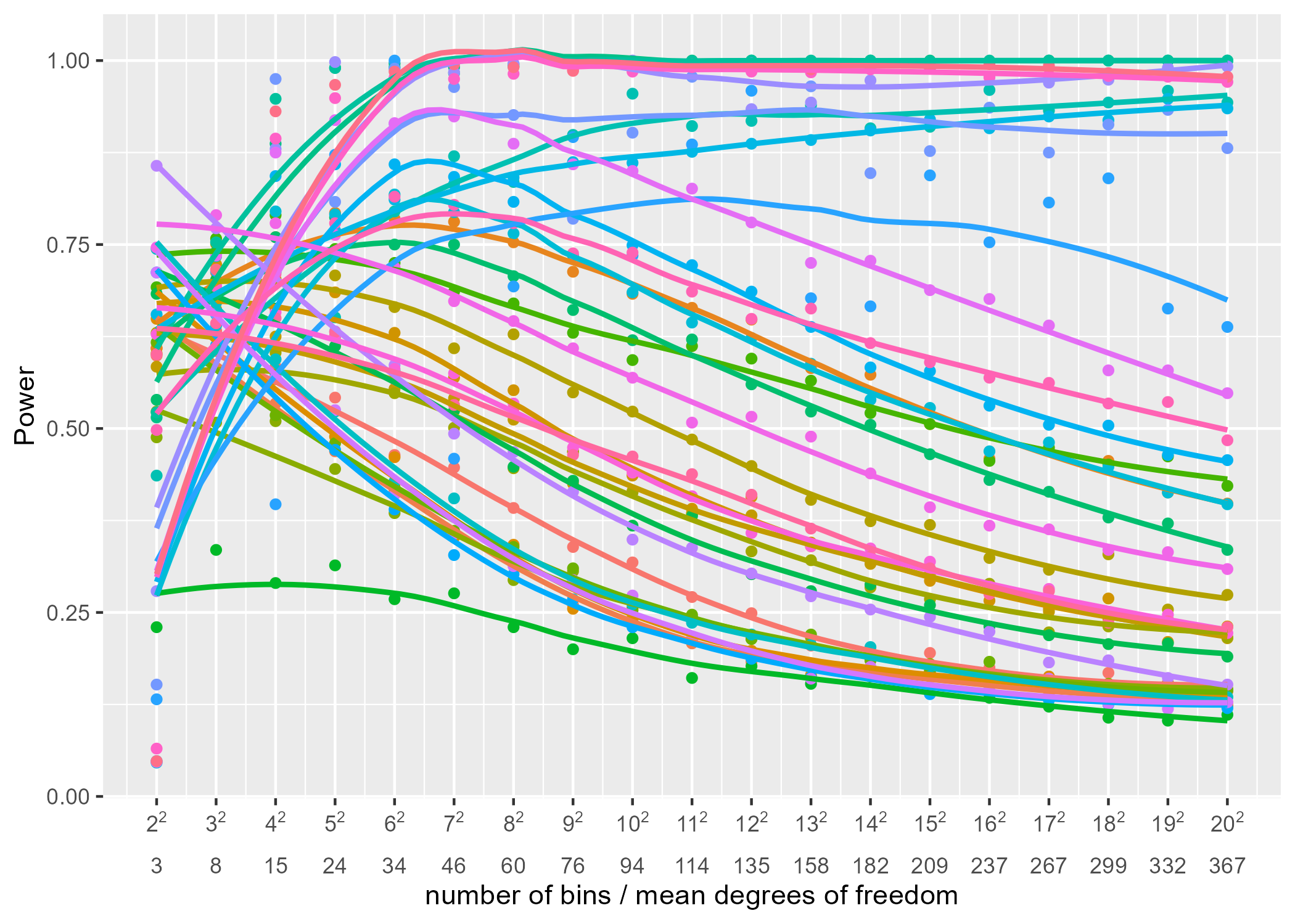}

\subsection{Goodness-of-fit methods for continuous data}

The goodness-of-fit problem is one of the oldest problems in Statistics, and there are a plethora of methods for univeriate data. Surprisingly, this is not true for multivariate data, where there are only a very few methods that have been discussed in the literature. To be precise, we are looking for methods with the following features:

\begin{enumerate}
\item
 The method should be a true omnibus test, that is it should not be designed for a specific distribution (like the multivariate normal) either under the null hypothesis or the alternative.  
\item
 The method should allow for parameter estimation.  
\item
 The method should not require the generation of an MC data set and run a subsequent two-sample test. 
\item
 we do not require a large sample theory but can always find p-values via simulation.  
\end{enumerate}

To understand why there are so few such methods, let us see what it would mean to extend one of the classic univariate goodness-of-fit tests to multivariate data, namely the Kolmogorov-Smirnov test. It is based on the largest absolute deviation of the cumulative and the empirical distribution functions:

$$\psi(F,\hat{F})=\max\left\{\vert F(\mathbf{x})-\hat{F}(\mathbf{x}\vert:\mathbf{x} \in \mathbf{R^d}\right\}$$

Now in principle this maximum is taken over all of $\mathbf{R^d}$, however in one dimension it is easy to see that the maximum will always be obtained at one of the data points, and therefore can be found with $n$ evaluations of $F$. Unfortunately in higher dimensions this is no longer true. Here the maximum can occur at any point who's coordinates are any combination of the coordinates of the points in the data set, and there are $n^d$ such points. So even for modest $n$ and $d$ this is impossible.

Some attempts have been made to find short-cuts, for example the Fasano-Franceschini test described later. In \emph{MDgof} we have implemented a very simple version, where (as in one dimension) we just find the maximum among the data points. Clearly one would expect this method to have lower power than a ``full'' test, but at least it can be applied to reasonably
large data sets. Several other methods also based on a comparison of the theoretical and the empirical distribution function have also been implemented in the same manner. As far as we know, these tests have not been discussed in the literature for data with more than one dimension, although they are easy to implement and can have good power, as our simulation studies have shown.  They are:

\subsubsection{Kolmogorov-Smirnov test (qKS)}

$$TS=\max\left\{\vert F(\mathbf{x}_i)-\hat{F}(\mathbf{x}_i\vert\right\}$$

The KS test was first proposed in \cite{Kolmogorov1933} and \cite{Smirnov1939}. We use the notation \emph{qKS} (quick Kolmogorov-Smirnov) to distinguish the test implemented in \emph{MDgof} from the full test. 

\subsubsection{Kuiper's test (qK)}

This is a variation of the KS test proposed in \cite{Kuiper1960}:

$$\psi(F,\hat{F})=\max\left\{ F(\mathbf{x})-\hat{F}(\mathbf{x}):\mathbf{x} \in \mathbf{R^d}\right\}+\max\left\{\hat{F}(\mathbf{x})-F(\mathbf{x}):\mathbf{x}  \in \mathbf{R^d}\right\}$$
$$TS=\max\left\{ F(\mathbf{x}_i)-\hat{F}(\mathbf{x}_i)\right\}+\max\left\{\hat{F}(\mathbf{x}_i)-F(\mathbf{x}_i)\right\}$$

\subsubsection{Cramer-vonMises test (qCvM)}

Another classic test based on

$$\psi(F,\hat{F})=\int \left(F(\mathbf{x})-\hat{F}(\mathbf{x})\right)^2 d\mathbf{x}$$

is implemented with 

$$TS=\sum_{i=1}^n \left(F(\mathbf{x}_i)-\hat{F}(\mathbf{x}_i)\right)^2$$

This test was first discussed in \cite{Anderson1962}.

\subsubsection{Anderson-Darling test (qAD)}

The Anderson-Darling test is based on

$$\psi(F,\hat{F})=\int \frac{\left(F(\mathbf{x})-\hat{F}(\mathbf{x})\right)^2}{F(\mathbf{x})[1-F(\mathbf{x})]} d\mathbf{x}$$

We use the test statistic

$$TS=\sum_{i=1}^n \frac{\left(F(\mathbf{x}_i)-\hat{F}(\mathbf{x}_i)\right)^2}{F(\mathbf{x}_i)[1-F(\mathbf{x}_i)]}$$

It was first proposed in \cite{anderson1952}.

In the following we discuss the few tests that have been proposed in the literature:

\subsubsection{Bickel-Breiman test (BB)}

This test uses the density, not the cumulative distribution function.

Let $R_j=\min \left\{||\mathbf{x}_i-\mathbf{x}_j||:1\le i\ne j \le n\right\}$ be some distance measure in $\mathbf{R}^d$, not necessarily Euclidean distance. Let $f$ be the density function under the null hypothesis and define

$$U_j=\exp\left[ -n\int_{||\mathbf{x}-\mathbf{x}_i||<R_j}f(\mathbf{x})d\mathbf{x}\right]$$

Then it can be shown that under the null hypothesis $U_1,..,U_n$ have a uniform distribution on $[0,1]$, and a goodness-of-fit test for univariate data such as Kolmogorov-Smirnov can be applied. This test was first discussed in \cite{bickel1983}.

\subsubsection{Bakshaev-Rudzkis test (BR)}

This test proceeds by estimating the density via a kernel density estimator and then comparing it to the density specified in the null hypothesis. Details are discussed in \cite{bakshaev2015}.

\subsubsection{Rosenblatt transforms}

A number of tests have been proposed based on the Rosenblatt transform. This is a generalization of the probability integral transform. It transforms a random vector $(X_1,..,X_d)$ into $(U_1,..,U_d)$, where $U_i\sim U[0,1]$ and $U_i$ is independent of $U_j$. It works as follows:

$$
\begin{aligned}
&U_1  = F_{X_1}(x_1)\\
&U_2  = F_{X_2|X_1}(x_2|x_1)\\
&... \\
&U_d  = F_{X_d|X_1,..,X_{d-1}}(x_d|x_1,..,x_{d-1})\\
\end{aligned}
$$

and so requires knowledge of the conditional distribution functions. In our case of a goodness-of-fit test, however, these will generally not be know. One can show, though, that

$$
\begin{aligned}
&F_{X_1}(x_1)    = F(x_1, \infty,..,\infty)\\
&F_{X_2|X_1}(x_2|x_1)    = \frac{\frac{d}{dx_1}F(x_1, x_2,\infty,..,\infty)}{\frac{d}{dx_1}F(x_1, \infty,..\infty)}\\
&... \\
&F_{X_d|X_1,..,X_{d-1}}(x_d|x_1,..,x_{d-1})    = \frac{\frac{d^{d-1}}{dx_1x_2..x_{d-1}}F(x_1,.., x_d)}{\frac{d^{d-1}}{dx_1x_2..x_{d-1}}F(x_1, \infty,..,\infty)}\\
\end{aligned}
$$

Unfortunately for general cdf $F$, these derivatives will have to be found numerically, and for $d>2$ this would not be feasible because of issues with calculation times and numerical instabilities. For these reasons these methods are only implemented for bivariate data.

\emph{MDgof} includes two tests based on the Rosenblatt transform:

\subsubsection{Fasano-Franceschini test (FF)}

This implements a simplified version of the Kolmogorov-Smirnov test after a Rosenblatt transform. It is discussed in \cite{Fasano1987}.

\subsubsection{Ripley's K test (RK)}

This test finds the number of observations with a radius r of a given observation for different values of R. After the Rosenblatt transform (if the null hypothesis is true) the data is supposed to be independent uniforms, and so the area of a circle of radius r is $\pi r^2$. The two are the compared via the mean square. This test was proposed in \cite{ripley1976}. The test is implemented in \emph{MDgof} using the R library \emph{spatstat} \cite{baddeley2005}.

\subsection{Goodness-of-fit methods for discrete data}

In the case of discrete (as well as histogram) data methods based on comparing the theoretical and the empirical distribution function actually work again as in one dimension because now the points are known and it is the counts that are random. For this reason the implementations of the Kolmorogov-Smrinov test (KS), Kuiper's test (K), Cramer-vonMises test (CvM) and Anderson-Darling test (AD) are straight forward and are no longer short-cuts. 

In addition \emph{MDgof} has a number of tests that similarly to the chi-square test compare the observed and the expected counts. They are

\subsubsection{Pearson's chi-square (P)}

$$\sum_{ij} \frac{(O_{ij}-E_{ij})^2}{E_{ij}}$$

\subsubsection{Total Variation (TV)}

$$\frac1{n^2}\sum_{ij} \left(O_{ij}-E_{ij}\right)^2$$

\subsubsection{Kullback-Leibler (KL)}

$$\frac1{n}\sum_{ij} O_{ij}\log\left(O_{ij}/E_{ij}\right)$$

\subsubsection{Hellinger (H)}

$$\frac1{n}\sum_{ij} \left(\sqrt{O_{ij}}-\sqrt{E_{ij}}\right)^2$$

\subsection{Two-sample methods for continuous data}

\subsubsection{Methods based on empirical distribution functions}

These are the classic methods Kolmogorov-Smirnon (KS) (Kolmogorov 1933) and (Smirnov 1939), Kuiper (K) (Kuiper 1960), Cramer-vonMises (CvM) (T. W. Anderson 1962) and Anderson-Darling (AD) (Theodore W. Anderson, Darling, et al. 1952). Unlike in the goodness-of-fit case here we compare two discrete functions, and so these are full tests, not shortcuts. 

\subsubsection{Methods based on nearest neighbors}

The test statistics are the average number of nearest neighbors of the
\(\mathbf{x}\) data set that are also from \(\mathbf{x}\) plus the
average number of nearest neighbors of the \(\mathbf{y}\) data set that
are also from \(\mathbf{y}\). \emph{NN1} uses one nearest neighbor and
\(NN5\) uses 5.

Next we have a number of methods based on distances between observations. We denote by \(||.||\) Euclidean distance.

\subsubsection{Aslan-Zech test (AZ)}

This test discussed in (Aslan and and 2005) uses the test statistic

\[
\begin{aligned}
&\frac{1}{nm}\sum_{i=1}^n \sum_{j=1}^m \log(||x_i-y_j||) -\\
&\frac{1}{n^2}\sum_{i=1}^n \sum_{i<j} \log(||x_i-x_j||) - \\
&\frac{1}{m^2}\sum_{i=1}^m \sum_{i<j} \log(||y_i-y_j||) 
\end{aligned}
\] 

\subsubsection{Baringhaus-Franz test (BF)}

Similar to the Aslan-Zech test, it uses the test statistic

\[
\begin{aligned}
&\frac{nm}{n+m}\left[\frac{1}{nm}\sum_{i=1}^n \sum_{j=1}^m \sqrt{||x_i-y_j||} + \right.\\
&\frac{1}{n^2}\sum_{i=1}^n \sum_{i<j} \sqrt{||x_i-x_j||} -\\
&\left. \frac{1}{m^2}\sum_{i=1}^m \sum_{i<j} \sqrt{||y_i-y_j||} \right]\\
\end{aligned}
\]

 and was first proposed in (Baringhaus and Franz 2004).

\subsubsection{Biswas-Ghosh test (BG)}

Another variation of a test based on Euclidean distance was discussed in
(Biswas and Ghosh 2014).

\[
\begin{aligned}
&B_{xy} = \frac{1}{nm}\sum_{i=1}^n \sum_{j=1}^m \sqrt{||x_i-y_j||} \\
&B_{xx}= \frac{2}{n(n-1)}\sum_{i=1}^n \sum_{i<j} \sqrt{||x_i-x_j||} \\
&B_{yy}=\frac{2}{m(m-1)}\sum_{i=1}^m \sum_{i<j} \sqrt{||y_i-y_j||}\\
&\left(B_{xx}-B_{xy}\right)^2+\left(B_{yy}-B_{xy}\right)^2
\end{aligned}
\]

\subsubsection{Maximum Mean Discrepancy test (MMD)}

The Maximum Mean Discrepancy (MMD) two-sample test is a nonparametric kernel-based test. It works by embedding both the kernel density estimates of the two data sets into a reproducing kernel Hilbert space (RKHS) and measuring the distance between their mean embeddings. If this distance (the MMD) is zero, the two distributions are identical for characteristic kernels; larger values indicate a discrepancy. The aggregated version implemented in \emph{MD2sample} uses 10 different bandwidths for the kernel density estimators. For detailed discussions see \cite{gretton2012B}, \cite{chwialkowski2016} and \cite{gretton2023}.

The following methods find p values using a large sample theory.

\subsubsection{Friedman-Rafski test (FR)}

This test is a multi-dimensional extension of the classic Wald-Wolfowitz
test bases on minimum spanning trees. It was discussed in (Friedman and
Rafsky 1979).

\subsubsection{Simple Nearest Neighbor test (NN)}

Similar to the nearest neigboor tests described earlier, it uses only
the number of nearest neighbors of the first data set that are also from
the first data set. This number has a binomial distribution, and this
can be used to find p values.

\subsubsection{Chen-Friedman tests (CF1 to CF4)}

These tests, discussed in (Chen and Friedman, n.d.), are implemented in
the \emph{gTests} (Chen and Zhang 2017) package.

\subsubsection{Ball Divergence test (Ball)}

A test described in (Pan et al. 2018) and implemented in the R package \emph{Ball}
(Zhu et al. 2021).

\subsection{Two-sample Problem - discrete data}

Implemented for discrete data are versions of the Kolmogorov-Smirnov (KS),
Kuiper (K), Cramer-vonMises (CvM), Anderson-Dar
ling (AD), nearest neighboor (NN),
Aslan-Zech (AZ), Baringhaus-Franz (BF) tests as well as a chisquare test.

\section{Case studies}

The conclusions described in this paper regarding the powers of the various methods are based on numerous case studies. We have the following different types of studies:

\begin{enumerate}
\item
 goodness-of-fit or two-sample problem.    
\item
 continuous or discrete data.  
\item
 2 or 5 dimensional data.  
\item
 studies with marginals that have the same distribution, or not.  
\item
 studies with or without parameter estimation (for goodness-of-fit problem).  
\end{enumerate}

For the goodness-of-fit problem there are 30 case studies each for cases with and without parameter estimation in two dimensions plus 30 cases for five dimensional data.  Half of each of these have equal marginals and half do not. For the two-sample problem there are 50 cases studies. The discrete cases are discretized versions of the continuous cases. Both \emph{MDgof} and \emph{MD2sample} include vignettes that describe all the case studies in detail. Here we will only show a few select examples.

In these examples the difference between a data set generated under the null hypothesis and one generated under the alternative is quite obvious for the purpose of illustration. In the actual case studies this is generally not true, so an actual test is needed. 

The alternative distribution always depends on a parameter that the user can choose. In the results described in this paper there are always two of these. One is chosen so that the null hypothesis is actually true (and one can check that the methods achieve the nominal type I error) and a second one is chosen in such a way that for the selected sample size and a true type I error of $\alpha=0.05$ the best method has a power around $80-90\%$. The sample sizes are $250$ for the goodness-of-fit studies and $200$ for the two-sample studies.

The following is worth noting: Say we wish to determine whether method A or method B has a higher power for a specific case (aka $X\sim F$ vs $X\sim G(p)$). We run a power study using $\alpha=0.05, p=1$ and a sample size of $n=100$ (say), and for this combination method A has a higher power. Than (almost always) A will also be the better method for a combination $\alpha=0.01, p=0.5, n=300$ or indeed any other combination. For this reason for any one specific case it suffices to run just one power study to find which method is best, second best and so on. 

\subsection{Naming conventions}

We have attempted to name the case studies in a manner that let's the user make an educated guess as to the nature of the study:

\subsubsection{Goodness-of-fit} 

A typical case study is called dist1.dist2, where dist1 is the distribution under the null hypothesis and dist2 is the true distribution of the data. For example, the third study is called normal-ind.normal-cor, indicating that under the null hypothesis the data comes from independent normal distributions whereas the true data comes from correlated normal distributions.

If the case study has distributions that have different marginals the name ends in .marginal. 

\subsubsection{Two-sample}

Here a typical name is dist, indicating that the two data sets come from distribution dist with different parameters. All names end in

\begin{itemize}
\item  D2: two dimensions, equal marginals  
\item  M:  two dimensions, unequal marginals  
\item  D5: five dimensions, equal marginals  
\item  M5:  five dimensions, unequal marginals  
\end{itemize}
 
\subsection{Examples}

\subsubsection{Example 1}

The first case study for the goodness-of-fit problem with continuous data, equal marginals, D=2 and no parameter estimation has under the null hypothesis independent uniform distributions in both the x and y variable. Under the alternative hypothesis (or the true data) a horizontal stripe is added to the data in such a way that the marginal distributions are still uniform, and so any univariate goodness-of-fit test would be unable to detect any difference and correctly reject the null hypothesis. An example is shown in figures 1 (scatterplots of data sets under the null and under the alternative) and 2 (histogram of x under the alternative, showing these are still uniform):

\includegraphics{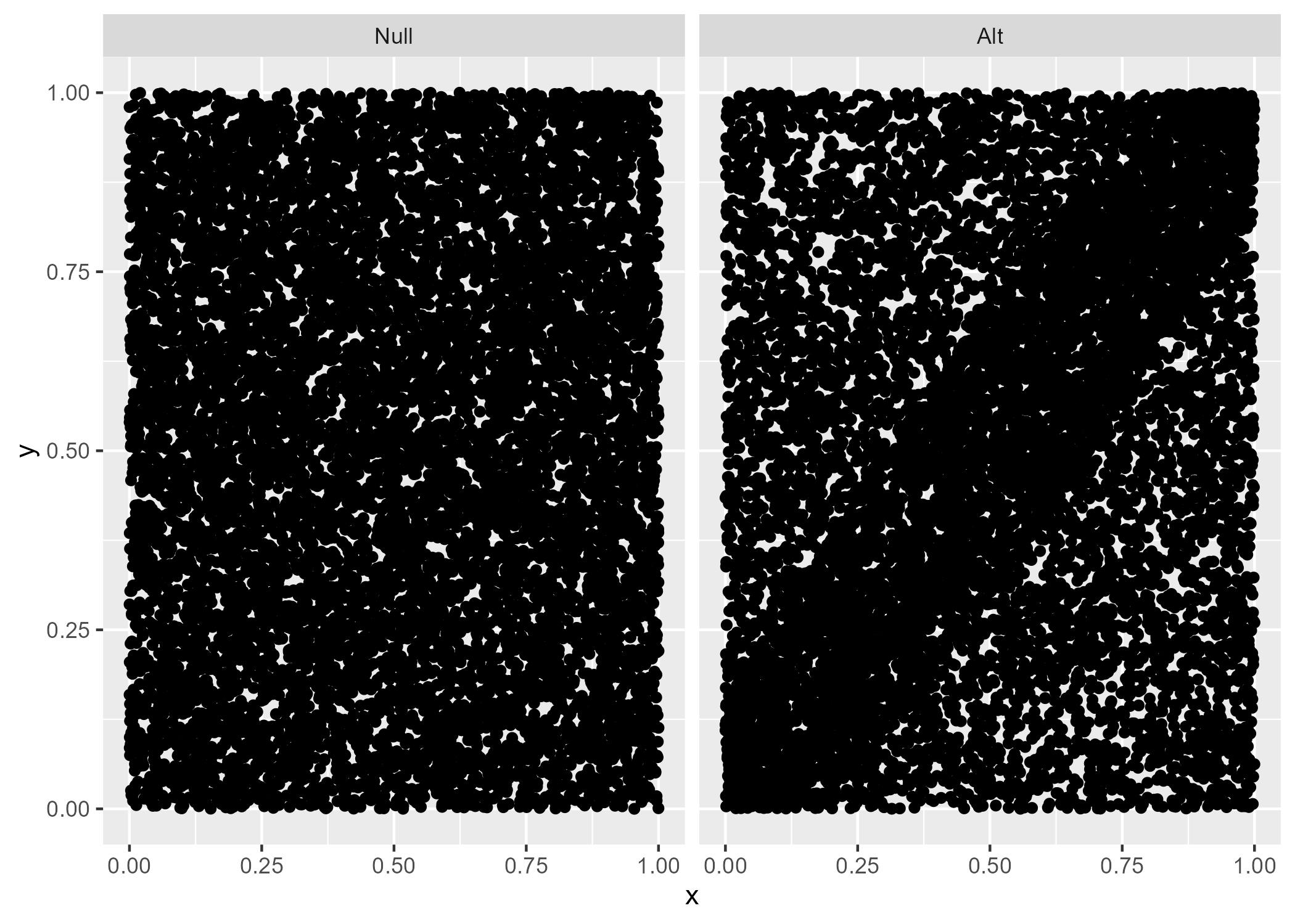}

\includegraphics{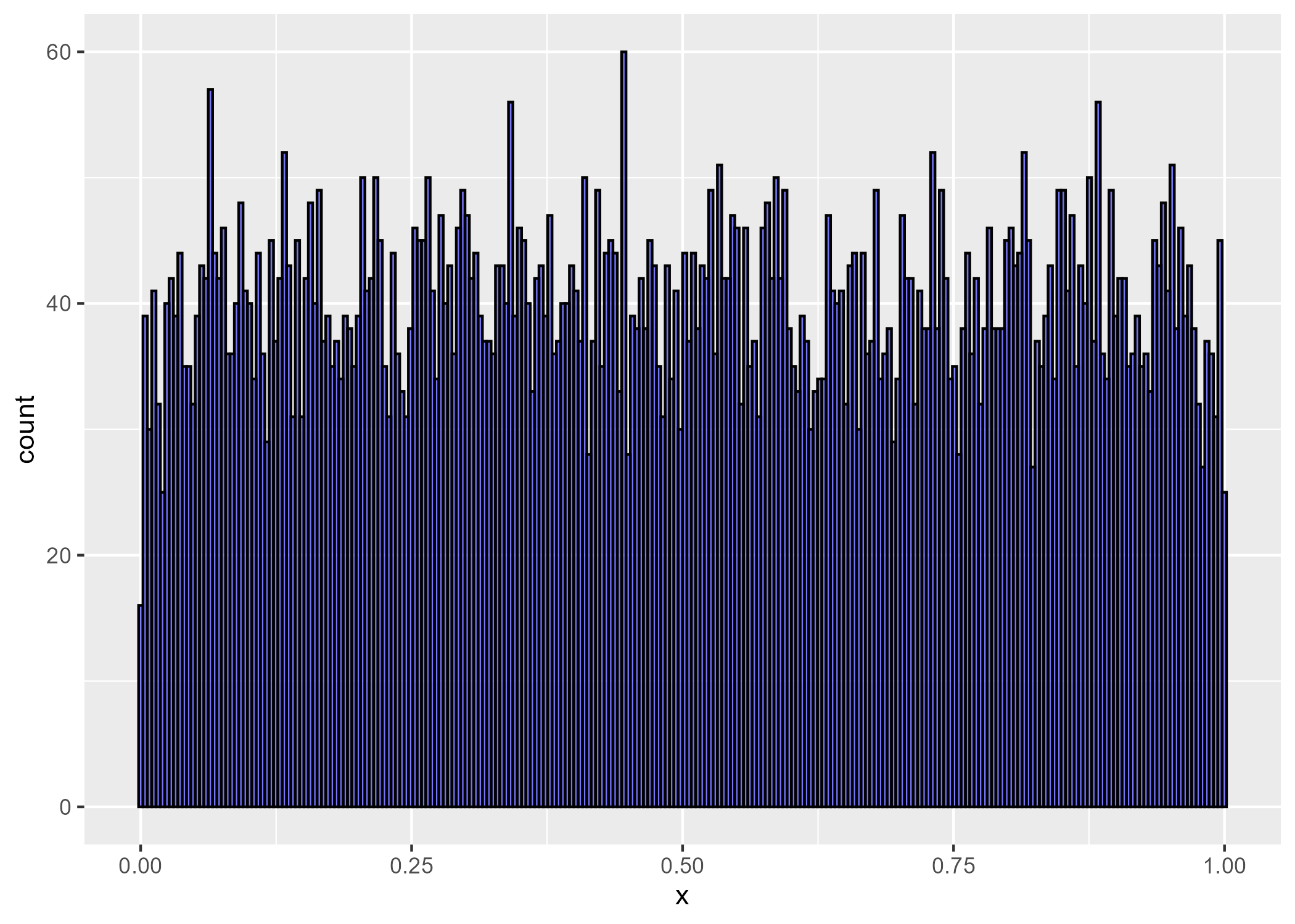}

The results for this case study is recorded in the file MDgof::power\textunderscore studies\textunderscore cont\textunderscore results, included in the $\mathbf{R}$ package \emph{MDgof}. Generally these files are lists with three elements:

\begin{enumerate}
\item
 Null - the powers when the null hypothesis is true   
\item
 Pow - the powers when the null hypothesis is false   
\item
 param\textunderscore alt - the parameter values under the alternative used   
\end{enumerate}

For the first case study we find the powers in table \ref{tblEx1}.

\input{tables/tblEx1.tex}

so the chi-squared tests are best with a power about $95\%$, followed by the quick Anderson-Darling method with a power of about $69\%$. Of course in the case of uniform distributions there is no difference between equal-size and equal-probability bins.

\subsubsection{Example 2}

As an example for discrete data, we use case 15. As with all the case studies for discrete data, these begin with the corresponding continuous case study and then bin the data into 5 by 5 grid. Case study 15 has as the null distribution a 50-50 mixture of Joe and Frank copulas, with the alternative an $(\lambda, 1-\lambda)$ mixture. Figure 3 shows an example.

\includegraphics{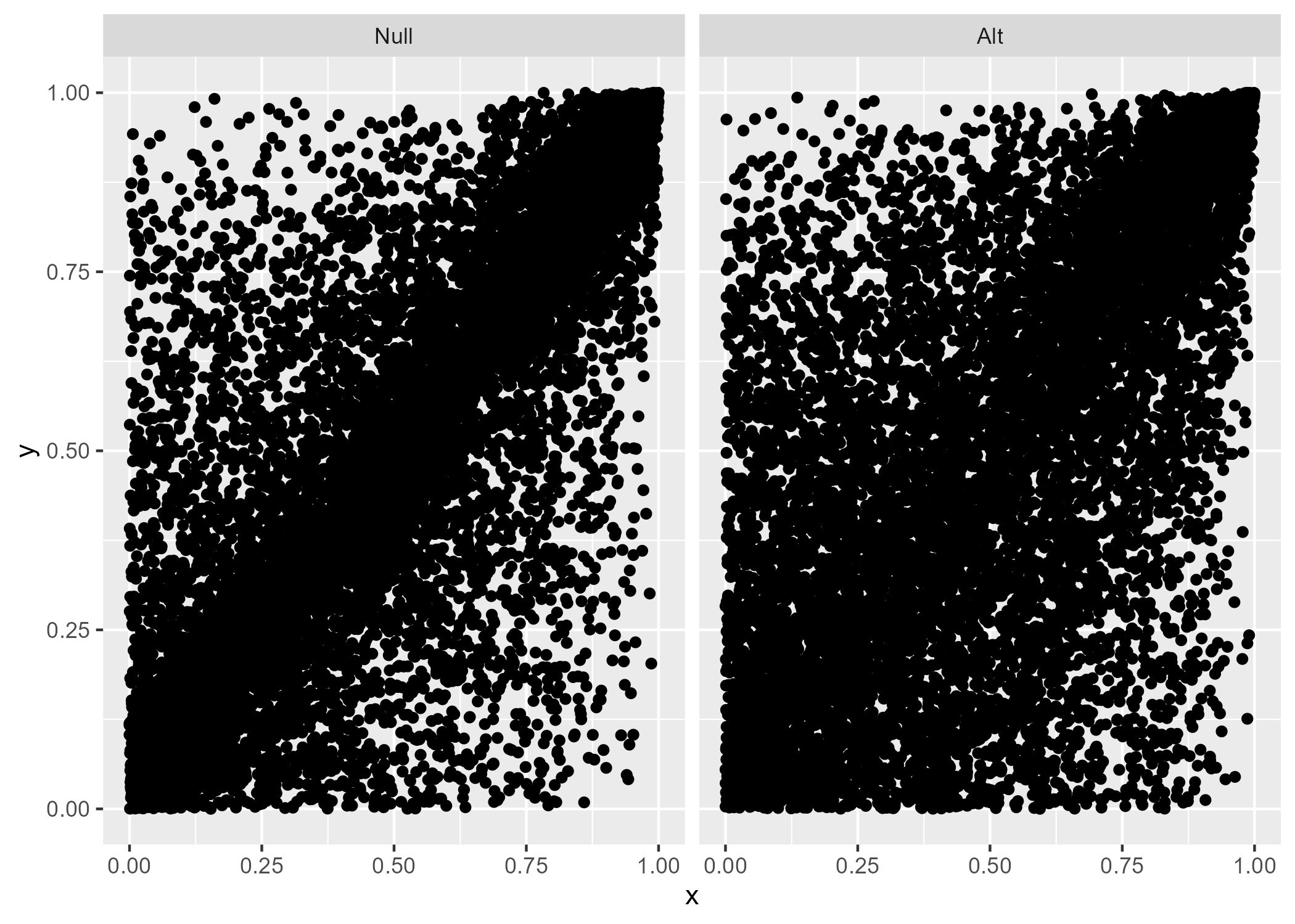}

Table \ref{tblDataEx2} has an example data set. The first two columns give the grid points and the third the counts.

\input{tables/tblDataEx2.tex}

The results for this case study is recorded in the file MDgof::power\textunderscore studies\textunderscore disc\textunderscore results. Table \ref{tblEx2} has the powers.

\input{tables/tblEx2.tex}

\subsubsection{Example 3}

As an example of a two-sample problem we use case study 15. This is data from a Dalitz plot, a type of two dimensional data encountered in high energy physics. Under the null hypothesis we have a uniform distribution on an area in the unit square bounded by quadratic functions. Under the alternative there might be a number of additional features which would be indications of new physics. An example is shown in figure 4.

\includegraphics{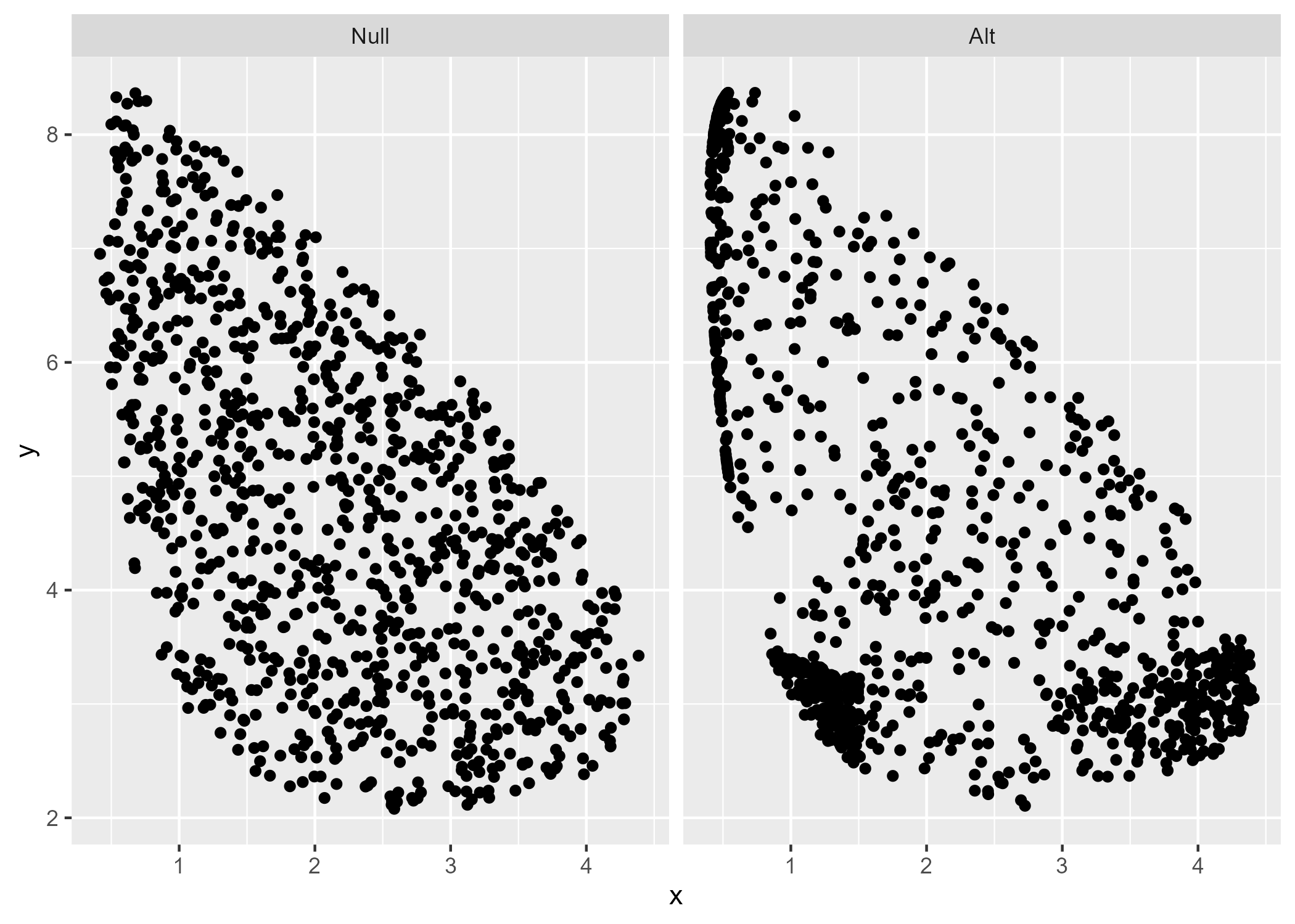}

The results of the power study are in  table \ref{tblEx3}. Here the method by Biswas and Ghosh is best, with a power of $94\%$, followed by Alsan and Zech ($89\%$).

\input{tables/tblEx3.tex}

The results of the power studies for the two-sample problem are recorded in the file MD2sample::power\textunderscore studies\textunderscore results. This file, part of \emph{MD2sample}, is a list with six elements:

\begin{itemize}
\item \emph{Null Cont}: true type I error for continuous data
\item \emph{Alt Cont}: power estimates for continuous data
\item \emph{Par Cont}: parameter values for alternative and continuous data
\item \emph{Null Disc}: true type I error for discrete data
\item \emph{Alt Disc}: power estimates for discrete data
\item \emph{Par Disc}: parameter values for alternative and discrete data
\end{itemize}

\newpage

\section{Results}

We will now give the results of the various studies. For each we present the following three tables:

\begin{enumerate}
\item
Mean power: the mean power of each method over the included case studies.  
\item
Percentage close to best: the number of times a method had a power close (aka within $90\%$) of the best method.  
\item
Best selection(s): for the combination of all case studies we find the smallest selection of methods such that for each study at least one of the methods is best. Usually there is more than one such selection. 
\end{enumerate}

\subsection{Goodness-of-fit, overall results}

In this section we include all case studies, whether they have estimated parameters or not and equal marginals or not

\subsubsection{Continuous methods - Dim=2}

\input{tables/tblGC2.tex}

The equal-size and equal-probability chi-square tests with  5 by 5 grid are best, followed by Fasano-Franceschini. The best selection requires seven methods, all including BR, EP, FF, qAD, qK and RK plus either qKS or qCvM.

\subsubsection{Continuous methods - Dim=5}

\input{tables/tblGC5.tex}

\subsubsection{Discrete methods}

\input{tables/tblGD2.tex}

\subsubsection{Continuous plus Hybrid methods}

To distinguish between methods that exist for both the goodness-of-fit and two-sample problem the abbreviations of the first end in G and the second in T1 if the MC sample size is equal to the one of the data set and T5 if it is five times as large.

\input{tables/tblHybrid1.tex}

\input{tables/tblHybrid1D5.tex}

\subsubsection{Continuous methods, with or without equal marginals}

Here we distinguish between case studies with marginals that are equal under the null and the alternative (so that univariate methods would not detect a difference) and those with different marginals.

1. With equal marginals

\input{tables/tblGCeM.tex}

2. Without equal marginals

\input{tables/tblGCuM.tex}

There is a remarkable difference between these two types of case studies. In the case of equal marginals only three methods are needed for the best selection, EP, qAD and RK. In the case of unequal marginals one needs six, and there are selections that do not include EP. Furthermore in the case of equal marginals the chi-square tests are much better than any others, with a mean power of $85\%$ for EP. 

\subsection{Two-sample}

\subsubsection{Continuous methods - overall results}

\input{tables/tblTC2.tex}

\input{tables/tblTC5.tex}

In both 2 and 5 dimensions MMD is best, followed by AZ.

\subsection{Continuous methods - with or without equal marginals}

1. With equal marginals

\input{tables/tblTCM.tex}

2. Without equal marginals

\input{tables/tblTCnM.tex}

\subsubsection{Discrete methods - overall results}

\input{tables/tblTD.tex}

\subsection{Discrete methods - with or without equal marginals}

1. With equal marginals

\input{tables/tblTDM.tex}

2. Without equal marginals

\input{tables/tblTDnM.tex}

\section{Continuous vs Discrete Methods}

In many of the case studies we ran both continuous and discrete methods on the same data sets, after discretizing the continuous data set. This allows us to compare methods from both types of data. The result is shown in figure 5. We see that, as one would expect, mostly discretizing comes at the cost of some power. However, there are some exceptions, especially on the goodness-of-fit problems.

\includegraphics{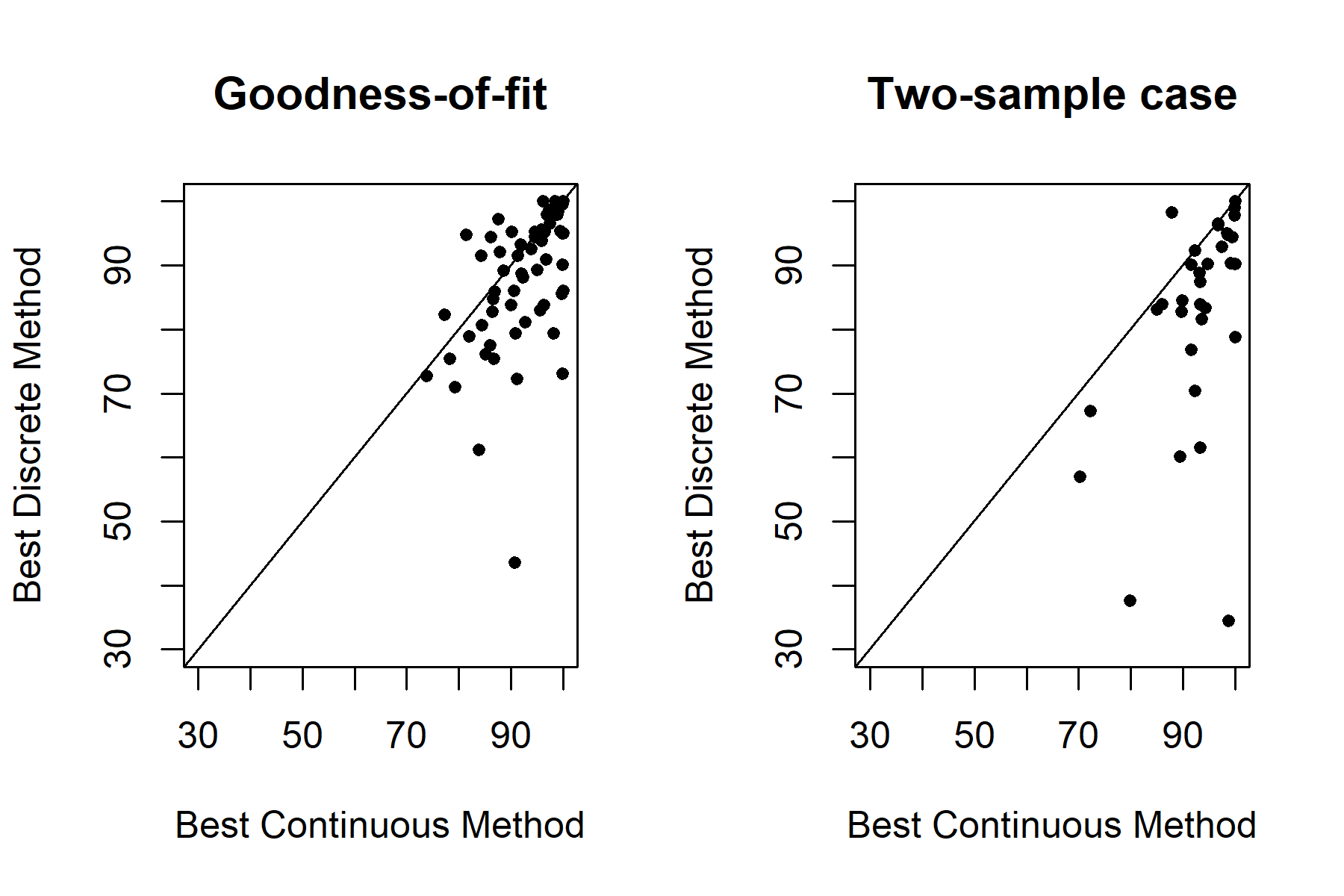}

\section{Detailed results}

In this section we show the detailed results of all the power studies. These are presented as follows. For each case study the methods are sorted according to their power, with the method with the highest power first. In the tables these are shown in dark red. Next come any method whose power is at least $95\%$ of the highest, essentially within simulation error, and these are shown in light red. Any method with a power at least $75\%$ of best is in green and within $50\%$ in blue. Any method with a power less than $50\%$ of best is shown without color.

The actual powers, as well as the actual type I errors, are included as files in the R packages.

\subsection{Goodness-of-fit}

\setlength{\tabcolsep}{5pt}   
\renewcommand{\arraystretch}{0.75}

\subsubsection{Continuous data, no estimation, D=2}

\input{tables/tblG.P.C2.nE.tex}

\subsubsection{Continuous data, with estimation, D=2}

\input{tables/tblG.P.C2.E.tex}

\subsubsection{Continuous data, no estimation, D=5}

\input{tables/tblG.P.C5.tex}

\subsubsection{Continuous data, hybrid methods, no estimation, nMC=1, Dim=2}

\input{tables/tblG.P.H22.tex}

\subsubsection{Continuous data, hybrid methods, no estimation, nMC=5, Dim=2}

\input{tables/tblG.P.H52.tex}

\subsubsection{Continuous data, hybrid methods, no estimation, nMC=1, Dim=5}

\input{tables/tblG.P.H25.tex}

\subsubsection{Continuous data, hybrid methods, no estimation, nMC=5, Dim=5}

\input{tables/tblG.P.H55.tex}

\subsubsection{Continuous data, gof+hybrid methods, no estimation, Dim=2}

\input{tables/tblGH.P.2.tex}

\subsubsection{Continuous data, gof+hybrid methods, no estimation, Dim=5}

\input{tables/tblGH.P.5.tex}

\subsubsection{Discrete data, no estimation}

\input{tables/tblG.P.D.nE.tex}

\subsubsection{Discrete data, with estimation}

\input{tables/tblG.P.D.E.tex}

\subsection{Two-sample}

\subsubsection{Continuous data, D=2}

\input{tables/tblT.P.C2.tex}

\subsubsection{Continuous data, D=5}

\input{tables/tblT.P.C5.tex}

\subsubsection{Discrete data}

\input{tables/tblT.P.D.tex}

\bibliographystyle{apalike}
\bibliography{references}

\end{document}

%% file: tables/tblEx1.tex
\begin{table}[H]
\centering
\begin{tabular}{lllll}
  \hline
  \hline
EP & ES & qAD & FF & qCvM \\ 
  96 & 95 & 69 & 62 & 57 \\ 
  ----- & ----- & ----- & ----- & ----- \\ 
  KSD & BR & RK & qKS & qK \\ 
  51 & 50 & 41 & 40 & 29 \\ 
  ----- & ----- & ----- & ----- & ----- \\ 
  BB &  &  &  &  \\ 
  8 &  &  &  &  \\ 
   \hline
\end{tabular}
\caption{Powers of various methods for case study 1 (goodness-of-fit problem), continuous data in two dimensions, no parameter estimation} 
\label{tblEx1}
\end{table}

%% file: tables/tblDataEx2.tex
\begin{table}[H]
\centering
\begin{tabular}{lll}
  \hline
  \hline
x & y & counts \\ 
  0.2 & 0.2 & 31 \\ 
  0.4 & 0.2 & 5 \\ 
  0.6 & 0.2 & 10 \\ 
  0.8 & 0.2 & 3 \\ 
  1 & 0.2 & 0 \\ 
  0.2 & 0.4 & 7 \\ 
  0.4 & 0.4 & 28 \\ 
  0.6 & 0.4 & 9 \\ 
  0.8 & 0.4 & 6 \\ 
  1 & 0.4 & 1 \\ 
  0.2 & 0.6 & 7 \\ 
  0.4 & 0.6 & 5 \\ 
  0.6 & 0.6 & 21 \\ 
  0.8 & 0.6 & 7 \\ 
  1 & 0.6 & 5 \\ 
  0.2 & 0.8 & 1 \\ 
  0.4 & 0.8 & 2 \\ 
  0.6 & 0.8 & 11 \\ 
  0.8 & 0.8 & 27 \\ 
  1 & 0.8 & 10 \\ 
  0.2 & 1 & 1 \\ 
  0.4 & 1 & 1 \\ 
  0.6 & 1 & 5 \\ 
  0.8 & 1 & 8 \\ 
  1 & 1 & 39 \\ 
   \hline
\end{tabular}
\caption{Example data set for case study 15 (goodness-of-fit problem), discrete data, no parameter estimation} 
\label{tblDataEx2}
\end{table}

%% file: tables/tblEx2.tex
\begin{table}[H]
\centering
\begin{tabular}{llllllll}
  \hline
  \hline
P & TV & KL & H & K & AD & CvM & KS \\ 
  93 & 88 & 83 & 51 & 37 & 28 & 25 & 21 \\ 
   \hline
\end{tabular}
\caption{Powers of various methods for case study 15 (goodness-of-fit problem), discrete data, no parameter estimation} 
\label{tblEx2}
\end{table}

%% file: tables/tblEx3.tex
\begin{table}[H]
\centering
\begin{tabular}{lllllllll}
  \hline
  \hline
BG & MMD & AZ & ES & AD & Ball & BF & NN5 & EP \\ 
  92 & 89 & 84 & 72 & 71 & 65 & 61 & 59 & 59 \\ 
  ----- & ----- & ----- & ----- & ----- & ----- & ----- & ----- & ----- \\ 
  FR & CF1 & CF3 & NN1 & CvM & CF4 & NN0 & K & KS \\ 
  49 & 49 & 49 & 43 & 40 & 39 & 33 & 33 & 31 \\ 
  ----- & ----- & ----- & ----- & ----- & ----- & ----- & ----- & ----- \\ 
  CF2 &  &  &  &  &  &  &  &  \\ 
  29 &  &  &  &  &  &  &  &  \\ 
   \hline
\end{tabular}
\caption{Powers of various methods for case study 24 (two-sample problem), continuous data} 
\label{tblEx3}
\end{table}

%% file: tables/tblGC2.tex
\begin{table}[H]
\centering
\caption{Continuous data, D=2, with and without estimation, with and without equal marginals}
\label{tab:GC2}

\begin{tabular}{lllllll}
\toprule
\multicolumn{7}{l}{\textbf{Mean Power}} \\
\midrule
  \hline
  \hline
ES & EP & FF & qK & RK & qCvM \\ 
  79 & 76 & 62 & 53 & 51 & 49 \\ 
  ----- & ----- & ----- & ----- & ----- & ----- \\ 
  qKS & qAD & BR & KSD & BB &  \\ 
  48 & 46 & 36 & 30 & 19 &  \\ 
   \hline
\addlinespace
\multicolumn{7}{l}{\textbf{Percentage close to best}} \\
\midrule
  \hline
  \hline
ES & EP & RK & qK & FF & qKS \\ 
  57 & 48 & 38 & 28 & 22 & 20 \\ 
  ----- & ----- & ----- & ----- & ----- & ----- \\ 
  qAD & qCvM & BR & KSD & BB &  \\ 
  20 & 18 & 15 & 10 & 2 &  \\ 
   \hline
\addlinespace
\multicolumn{7}{l}{\textbf{Best selection(s)}} \\
\midrule
  \hline
  \hline
BR & EP & FF & qAD & qK & qKS & RK \\ 
  BR & EP & FF & qAD & qCvM & qK & RK \\ 
   \hline
\bottomrule
\end{tabular}
\end{table}

%% file: tables/tblGC5.tex
\begin{table}[H]
\centering
\caption{Continuous data, D=5, with and without equal marginals}
\label{tab:GC5}

\begin{tabular}{lllll}
\toprule
\multicolumn{5}{l}{\textbf{Mean Power}} \\
\midrule
  \hline
  \hline
qCvM & qKS & qK & BR & qAD \\ 
  69 & 66 & 62 & 43 & 37 \\ 
   \hline
\addlinespace
\multicolumn{5}{l}{\textbf{Percentage close to best}} \\
\midrule
  \hline
  \hline
qCvM & qK & qKS & qAD & BR \\ 
  60 & 57 & 43 & 17 & 17 \\ 
   \hline
\addlinespace
\multicolumn{5}{l}{\textbf{Best selection(s)}} \\
\midrule
  \hline
  \hline
BR & qAD & qCvM & qK \\ 
   \hline
\bottomrule
\end{tabular}
\end{table}

%% file: tables/tblGD2.tex
\begin{table}[H]
\centering
\caption{Discrete data, with and without estimation, with and without       equal marginals}
\label{tab:GD2}

\begin{tabular}{llll}
\toprule
\multicolumn{4}{l}{\textbf{Mean Power}} \\
\midrule
  \hline
  \hline
KL & H & TV & P \\ 
  77 & 71 & 71 & 67 \\ 
  ----- & ----- & ----- & ----- \\ 
  K & AD & CvM & KS \\ 
  53 & 52 & 44 & 44 \\ 
   \hline
\addlinespace
\multicolumn{4}{l}{\textbf{Percentage close to best}} \\
\midrule
  \hline
  \hline
KL & P & H & K \\ 
  58 & 48 & 45 & 37 \\ 
  ----- & ----- & ----- & ----- \\ 
  TV & AD & KS & CvM \\ 
  33 & 23 & 18 & 17 \\ 
   \hline
\addlinespace
\multicolumn{4}{l}{\textbf{Best selection(s)}} \\
\midrule
  \hline
  \hline
CvM & H & K & P \\ 
  CvM & K & KL & P \\ 
  AD & H & K & P \\ 
  AD & K & KL & P \\ 
   \hline
\bottomrule
\end{tabular}
\end{table}

%% file: tables/tblHybrid1.tex
\begin{table}[H]
\centering
\caption{Power of gof vs hybrid methods, Dim=2, methods with at least 50 mean power only}
\label{tab:Hybrid1}

\begin{tabular}{lllllllll}
\toprule
\multicolumn{9}{l}{\textbf{Mean Power}} \\
\midrule
  \hline
  \hline
ESG & EPG & MMDT5 & EST5 & MMDT1 & BFT5 & EPT5 & FFG & EST1 \\ 
  81 & 80 & 79 & 72 & 69 & 67 & 66 & 65 & 61 \\ 
  ----- & ----- & ----- & ----- & ----- & ----- & ----- & ----- & ----- \\ 
  BFT1 & qADG & qCvMG & EPT1 & ADT5 & qKG & qKSG & BallT5 & KT5 \\ 
  57 & 56 & 56 & 55 & 55 & 53 & 53 & 52 & 51 \\ 
   \hline
\addlinespace
\multicolumn{9}{l}{\textbf{Percentage close to best}} \\
\midrule
  \hline
  \hline
EPG & ESG & MMDT5 & EST5 & BallT5 & BFT5 & qKG & EPT5 & qKSG \\ 
  67 & 63 & 60 & 40 & 37 & 30 & 27 & 27 & 23 \\ 
  ----- & ----- & ----- & ----- & ----- & ----- & ----- & ----- & ----- \\ 
  qCvMG & FFG & qADG & MMDT1 & BFT1 & ADT5 & KT5 & EST1 & EPT1 \\ 
  20 & 20 & 17 & 17 & 13 & 10 & 3 & 0 & 0 \\ 
   \hline
\addlinespace
\multicolumn{9}{l}{\textbf{Best selection(s)}} \\
\midrule
  \hline
  \hline
BallT5 & EPG & KT5 & qKSG \\ 
  BallT5 & EPG & qKG & qKSG \\ 
  BallT5 & EPG & KT5 & qCvMG \\ 
  BallT5 & EPG & qCvMG & qKG \\ 
  BFT1 & EPG & qKG & qKSG \\ 
  BFT1 & EPG & qCvMG & qKG \\ 
  BallT5 & EPG & FFG & KT5 \\ 
  BallT5 & EPG & FFG & qKG \\ 
  BFT5 & EPG & qKG & qKSG \\ 
  BFT5 & EPG & qCvMG & qKG \\ 
  BallT5 & EPG & EST5 & qKSG \\ 
  BallT5 & EPG & EST5 & qCvMG \\ 
  BallT5 & EPG & MMDT5 & qKSG \\ 
  EPG & MMDT5 & qKG & qKSG \\ 
  BallT5 & EPG & MMDT5 & qCvMG \\ 
  EPG & MMDT5 & qCvMG & qKG \\ 
  BallT5 & EPG & EST5 & FFG \\ 
  BallT5 & BFT5 & EPG & EST5 \\ 
  BallT5 & EPG & FFG & MMDT5 \\ 
   \hline
\bottomrule
\end{tabular}
\end{table}

%% file: tables/tblHybrid1D5.tex
\begin{table}[H]
\centering
\caption{Power of gof vs hybrid methods, Dim=5, methods with at least 50 mean power only}
\label{tab:Hybrid1D5}

\begin{tabular}{lllllllll}
\toprule
\multicolumn{9}{l}{\textbf{Mean Power}} \\
\midrule
  \hline
  \hline
MMDT5 & AZT5 & MMDT1 & BFT5 & qCvMG & CvMT5 & KST5 & qKSG & BallT5 \\ 
  87 & 79 & 73 & 73 & 69 & 68 & 66 & 66 & 64 \\ 
  ----- & ----- & ----- & ----- & ----- & ----- & ----- & ----- & ----- \\ 
  AZT1 & qKG & ADT5 & KT5 & NN5T5 & BallT1 & BFT1 & CvMT1 & BGT5 \\ 
  63 & 62 & 60 & 59 & 57 & 56 & 55 & 53 & 52 \\ 
  ----- & ----- & ----- & ----- & ----- & ----- & ----- & ----- & ----- \\ 
  NN5T1 & ADT1 &  &  &  &  &  &  &  \\ 
  52 & 51 &  &  &  &  &  &  &  \\ 
   \hline
\addlinespace
\multicolumn{9}{l}{\textbf{Percentage close to best}} \\
\midrule
  \hline
  \hline
MMDT5 & AZT5 & BFT5 & BallT5 & MMDT1 & BallT1 & qCvMG & AZT1 & BFT1 \\ 
  80 & 60 & 57 & 47 & 43 & 43 & 40 & 37 & 37 \\ 
  ----- & ----- & ----- & ----- & ----- & ----- & ----- & ----- & ----- \\ 
  KT5 & BGT5 & KST5 & CvMT5 & qKSG & ADT5 & CvMT1 & ADT1 & NN5T5 \\ 
  37 & 37 & 33 & 27 & 23 & 20 & 17 & 17 & 17 \\ 
  ----- & ----- & ----- & ----- & ----- & ----- & ----- & ----- & ----- \\ 
  qKG & NN5T1 &  &  &  &  &  &  &  \\ 
  13 & 13 &  &  &  &  &  &  &  \\ 
   \hline
\addlinespace
\multicolumn{9}{l}{\textbf{Best selection(s)}} \\
\midrule
  \hline
  \hline
BGT5 & MMDT5 & NN5T5 & qCvMG \\ 
   \hline
\bottomrule
\end{tabular}
\end{table}

%% file: tables/tblGCeM.tex
\begin{table}[H]
\centering
\caption{Continuous data, with and without estimation, with equal marginals}
\label{tab:GCeM}

\begin{tabular}{llllll}
\toprule
\multicolumn{6}{l}{\textbf{Mean Power}} \\
\midrule
  \hline
  \hline
EP & ES & FF & RK & qAD & qCvM \\ 
  85 & 84 & 57 & 51 & 40 & 37 \\ 
  ----- & ----- & ----- & ----- & ----- & ----- \\ 
  qKS & BR & qK & BB & KSD &  \\ 
  30 & 27 & 26 & 25 & 23 &  \\ 
   \hline
\addlinespace
\multicolumn{6}{l}{\textbf{Percentage close to best}} \\
\midrule
  \hline
  \hline
EP & ES & RK & qAD & FF & qCvM \\ 
  73 & 70 & 40 & 10 & 10 & 7 \\ 
  ----- & ----- & ----- & ----- & ----- & ----- \\ 
  qKS & qK & BB & BR & KSD &  \\ 
  3 & 3 & 3 & 0 & 0 &  \\ 
   \hline
\addlinespace
\multicolumn{6}{l}{\textbf{Best selection(s)}} \\
\midrule
  \hline
  \hline
EP & qAD & RK \\ 
   \hline
\bottomrule
\end{tabular}
\end{table}

%% file: tables/tblGCuM.tex
\begin{table}[H]
\centering
\caption{Continuous data, with and without estimation, without equal marginals}
\label{tab:GCuM}

\begin{tabular}{llllll}
\toprule
\multicolumn{6}{l}{\textbf{Mean Power}} \\
\midrule
  \hline
  \hline
qK & ES & EP & FF & qKS & qCvM \\ 
  79 & 74 & 68 & 66 & 65 & 60 \\ 
  ----- & ----- & ----- & ----- & ----- & ----- \\ 
  qAD & RK & BR & KSD & BB &  \\ 
  52 & 51 & 45 & 38 & 13 &  \\ 
   \hline
\addlinespace
\multicolumn{6}{l}{\textbf{Percentage close to best}} \\
\midrule
  \hline
  \hline
qK & ES & qKS & RK & FF & qCvM \\ 
  53 & 43 & 37 & 37 & 33 & 30 \\ 
  ----- & ----- & ----- & ----- & ----- & ----- \\ 
  qAD & BR & EP & KSD & BB &  \\ 
  30 & 30 & 23 & 20 & 0 &  \\ 
   \hline
\addlinespace
\multicolumn{6}{l}{\textbf{Best selection(s)}} \\
\midrule
  \hline
  \hline
BR & FF & qAD & qCvM & qK & RK \\ 
  BR & FF & qAD & qK & qKS & RK \\ 
   \hline
\bottomrule
\end{tabular}
\end{table}

%% file: tables/tblTC2.tex
\begin{table}[H]
\centering
\caption{Continuous data, D=2, with and without equal marginals}
\label{tab:TC2}

\begin{tabular}{llllllllll}
\toprule
\multicolumn{10}{l}{\textbf{Mean Power}} \\
\midrule
  \hline
  \hline
MMD & AZ & ES & BF & EP & AD & NN5 & CvM & Ball & FR \\ 
  90 & 87 & 79 & 77 & 75 & 66 & 63 & 59 & 57 & 56 \\ 
  ----- & ----- & ----- & ----- & ----- & ----- & ----- & ----- & ----- & ----- \\ 
  CF1 & CF3 & KS & K & CF4 & NN1 & CF2 & NN0 & BG &  \\ 
  56 & 56 & 54 & 50 & 49 & 47 & 42 & 38 & 28 &  \\ 
   \hline
\addlinespace
\multicolumn{10}{l}{\textbf{Percentage close to best}} \\
\midrule
  \hline
  \hline
MMD & AZ & ES & EP & BF & Ball & CvM & AD & NN5 & FR \\ 
  91 & 85 & 53 & 41 & 38 & 26 & 24 & 24 & 18 & 18 \\ 
  ----- & ----- & ----- & ----- & ----- & ----- & ----- & ----- & ----- & ----- \\ 
  CF1 & CF3 & KS & K & BG & CF4 & NN1 & CF2 & NN0 &  \\ 
  18 & 18 & 15 & 15 & 15 & 9 & 6 & 6 & 3 &  \\ 
   \hline
\addlinespace
\multicolumn{10}{l}{\textbf{Best selection(s)}} \\
\midrule
  \hline
  \hline
Ball & BG & MMD \\ 
  BG & CvM & MMD \\ 
  AD & BG & MMD \\ 
  AZ & BG & ES \\ 
  BF & BG & MMD \\ 
  AZ & BG & MMD \\ 
   \hline
\bottomrule
\end{tabular}
\end{table}

%% file: tables/tblTC5.tex
\begin{table}[H]
\centering
\caption{Continuous data, D=5, with and without equal marginals}
\label{tab:TC5}

\begin{tabular}{lllllllll}
\toprule
\multicolumn{9}{l}{\textbf{Mean Power}} \\
\midrule
  \hline
  \hline
MMD & AZ & BF & AD & Ball & KS & CvM & NN5 & K \\ 
  82 & 74 & 67 & 57 & 57 & 54 & 54 & 54 & 53 \\ 
  ----- & ----- & ----- & ----- & ----- & ----- & ----- & ----- & ----- \\ 
  CF4 & FR & CF1 & CF3 & CF2 & BG & NN1 & NN0 &  \\ 
  43 & 40 & 40 & 40 & 39 & 38 & 33 & 28 &  \\ 
   \hline
\addlinespace
\multicolumn{9}{l}{\textbf{Percentage close to best}} \\
\midrule
  \hline
  \hline
MMD & AZ & BF & Ball & KS & K & NN5 & BG & CvM \\ 
  75 & 69 & 56 & 25 & 19 & 19 & 19 & 19 & 12 \\ 
  ----- & ----- & ----- & ----- & ----- & ----- & ----- & ----- & ----- \\ 
  AD & FR & CF1 & CF2 & CF3 & CF4 & NN1 & NN0 &  \\ 
  6 & 6 & 6 & 6 & 6 & 6 & 0 & 0 &  \\ 
   \hline
\addlinespace
\multicolumn{9}{l}{\textbf{Best selection(s)}} \\
\midrule
  \hline
  \hline
BG & MMD & NN5 \\ 
   \hline
\bottomrule
\end{tabular}
\end{table}

%% file: tables/tblTCM.tex
\begin{table}[H]
\centering
\caption{Continuous data, with equal marginals}
\label{tab:TCM}

\begin{tabular}{llllllllll}
\toprule
\multicolumn{10}{l}{\textbf{Mean Power}} \\
\midrule
  \hline
  \hline
MMD & AZ & ES & EP & BF & NN5 & FR & CF1 & CF3 & AD \\ 
  92 & 88 & 88 & 83 & 77 & 69 & 63 & 63 & 63 & 61 \\ 
  ----- & ----- & ----- & ----- & ----- & ----- & ----- & ----- & ----- & ----- \\ 
  CvM & CF4 & NN1 & CF2 & KS & Ball & NN0 & K & BG &  \\ 
  57 & 56 & 53 & 49 & 47 & 45 & 43 & 40 & 15 &  \\ 
   \hline
\addlinespace
\multicolumn{10}{l}{\textbf{Percentage close to best}} \\
\midrule
  \hline
  \hline
MMD & AZ & ES & EP & NN5 & BF & FR & CF1 & CF3 & CvM \\ 
  100 & 82 & 76 & 65 & 29 & 29 & 29 & 29 & 29 & 18 \\ 
  ----- & ----- & ----- & ----- & ----- & ----- & ----- & ----- & ----- & ----- \\ 
  Ball & AD & CF4 & KS & K & NN1 & NN0 & CF2 & BG &  \\ 
  18 & 12 & 12 & 6 & 6 & 6 & 6 & 6 & 0 &  \\ 
   \hline
\addlinespace
\multicolumn{10}{l}{\textbf{Best selection(s)}} \\
\midrule
  \hline
  \hline
BG & MMD \\ 
  K & MMD \\ 
  MMD & NN0 \\ 
  Ball & MMD \\ 
  KS & MMD \\ 
  CF2 & MMD \\ 
  MMD & NN1 \\ 
  CF4 & MMD \\ 
  CvM & MMD \\ 
  AD & MMD \\ 
  FR & MMD \\ 
  CF1 & MMD \\ 
  CF3 & MMD \\ 
  MMD & NN5 \\ 
  BF & MMD \\ 
  EP & MMD \\ 
  AZ & ES \\ 
  ES & MMD \\ 
  AZ & MMD \\ 
   \hline
\bottomrule
\end{tabular}
\end{table}

%% file: tables/tblTCnM.tex
\begin{table}[H]
\centering
\caption{Continuous data, without equal marginals}
\label{tab:TCnM}

\begin{tabular}{llllllllll}
\toprule
\multicolumn{10}{l}{\textbf{Mean Power}} \\
\midrule
  \hline
  \hline
MMD & AZ & BF & AD & ES & Ball & EP & CvM & KS & K \\ 
  88 & 86 & 78 & 72 & 70 & 69 & 67 & 62 & 60 & 60 \\ 
  ----- & ----- & ----- & ----- & ----- & ----- & ----- & ----- & ----- & ----- \\ 
  NN5 & FR & CF1 & CF3 & CF4 & BG & NN1 & CF2 & NN0 &  \\ 
  57 & 49 & 49 & 49 & 43 & 41 & 40 & 36 & 33 &  \\ 
   \hline
\addlinespace
\multicolumn{10}{l}{\textbf{Percentage close to best}} \\
\midrule
  \hline
  \hline
AZ & MMD & BF & AD & Ball & CvM & BG & ES & KS & K \\ 
  88 & 82 & 47 & 35 & 35 & 29 & 29 & 29 & 24 & 24 \\ 
  ----- & ----- & ----- & ----- & ----- & ----- & ----- & ----- & ----- & ----- \\ 
  EP & NN1 & NN5 & FR & CF1 & CF2 & CF3 & CF4 & NN0 &  \\ 
  18 & 6 & 6 & 6 & 6 & 6 & 6 & 6 & 0 &  \\ 
   \hline
\addlinespace
\multicolumn{10}{l}{\textbf{Best selection(s)}} \\
\midrule
  \hline
  \hline
AZ & BG \\ 
   \hline
\bottomrule
\end{tabular}
\end{table}

%% file: tables/tblTD.tex
\begin{table}[H]
\centering
\caption{Discrete data, with and without equal marginals}
\label{tab:TD}

\begin{tabular}{llllllll}
\toprule
\multicolumn{8}{l}{\textbf{Mean Power}} \\
\midrule
  \hline
  \hline
Chisquare & NN & AD & K & KS & CvM & AZ & BF \\ 
  79 & 53 & 49 & 47 & 43 & 43 & 27 & 20 \\ 
   \hline
\addlinespace
\multicolumn{8}{l}{\textbf{Percentage close to best}} \\
\midrule
  \hline
  \hline
Chisquare & K & KS & AD & NN & AZ & CvM & BF \\ 
  85 & 21 & 15 & 15 & 15 & 15 & 9 & 6 \\ 
   \hline
\addlinespace
\multicolumn{8}{l}{\textbf{Best selection(s)}} \\
\midrule
  \hline
  \hline
AZ & BF & Chisquare & KS \\ 
  AZ & Chisquare & CvM & KS \\ 
  AZ & Chisquare & CvM & K \\ 
  AD & AZ & Chisquare & KS \\ 
  AD & AZ & Chisquare & K \\ 
   \hline
\bottomrule
\end{tabular}
\end{table}

%% file: tables/tblTDM.tex
\begin{table}[H]
\centering
\caption{Discrete data, with equal marginals}
\label{tab:TDM}

\begin{tabular}{llllllll}
\toprule
\multicolumn{8}{l}{\textbf{Mean Power}} \\
\midrule
  \hline
  \hline
Chisquare & NN & AD & CvM & KS & K & AZ & BF \\ 
  89 & 62 & 49 & 45 & 38 & 33 & 12 & 12 \\ 
   \hline
\addlinespace
\multicolumn{8}{l}{\textbf{Percentage close to best}} \\
\midrule
  \hline
  \hline
Chisquare & NN & AD & KS & K & CvM & AZ & BF \\ 
  100 & 29 & 12 & 6 & 6 & 6 & 6 & 6 \\ 
   \hline
\addlinespace
\multicolumn{8}{l}{\textbf{Best selection(s)}} \\
\midrule
  \hline
  \hline
BF & Chisquare \\ 
  AZ & Chisquare \\ 
  Chisquare & K \\ 
  Chisquare & KS \\ 
  Chisquare & CvM \\ 
  AD & Chisquare \\ 
  Chisquare & NN \\ 
   \hline
\bottomrule
\end{tabular}
\end{table}

%% file: tables/tblTDnM.tex
\begin{table}[H]
\centering
\caption{Discrete data, without equal marginals}
\label{tab:TDnM}

\begin{tabular}{llllllll}
\toprule
\multicolumn{8}{l}{\textbf{Mean Power}} \\
\midrule
  \hline
  \hline
Chisquare & K & AD & KS & NN & AZ & CvM & BF \\ 
  70 & 62 & 50 & 49 & 45 & 41 & 41 & 28 \\ 
   \hline
\addlinespace
\multicolumn{8}{l}{\textbf{Percentage close to best}} \\
\midrule
  \hline
  \hline
Chisquare & K & KS & AZ & AD & CvM & BF & NN \\ 
  71 & 35 & 24 & 24 & 18 & 12 & 6 & 0 \\ 
   \hline
\addlinespace
\multicolumn{8}{l}{\textbf{Best selection(s)}} \\
\midrule
  \hline
  \hline
AZ & BF & Chisquare & KS \\ 
  AZ & Chisquare & CvM & KS \\ 
  AD & AZ & Chisquare & KS \\ 
  AZ & Chisquare & CvM & K \\ 
  AD & AZ & Chisquare & K \\ 
   \hline
\bottomrule
\end{tabular}
\end{table}

%% file: tables/tblG.P.C2.nE.tex
\begin{table}[H]
\centering
\begin{tabular}{lllllll}
  \hline
  \hline
uniform.uniform-diagonal-n & \cellcolor{red!40}BB & \cellcolor{red!20}ES & \cellcolor{blue!40}qAD & \cellcolor{blue!40}FF & \cellcolor{blue!40}qCvM & \cellcolor{blue!40}KSD \\ 
  uniform.uniform-diagonal-b & \cellcolor{red!40}BB & \cellcolor{red!20}ES & \cellcolor{green!40}qAD & \cellcolor{blue!40}qCvM & \cellcolor{blue!40}FF & BR \\ 
  normal-ind.normal-cor & \cellcolor{red!40}RK & \cellcolor{red!20}EP & \cellcolor{green!40}qAD & \cellcolor{blue!40}FF & \cellcolor{blue!40}qCvM & \cellcolor{blue!40}KSD \\ 
  t-ind.t-cor & \cellcolor{red!40}RK & \cellcolor{green!40}qAD & \cellcolor{green!40}ES & \cellcolor{blue!40}qCvM & \cellcolor{blue!40}FF & \cellcolor{blue!40}qKS \\ 
  uniform.Frank & \cellcolor{red!40}BB & \cellcolor{red!20}ES & \cellcolor{green!40}qAD & \cellcolor{blue!40}qCvM & \cellcolor{blue!40}FF & \cellcolor{blue!40}BR \\ 
  uniform.Clayton & \cellcolor{red!40}BB & \cellcolor{red!20}EP & \cellcolor{red!20}qAD & \cellcolor{blue!40}FF & \cellcolor{blue!40}BR & \cellcolor{blue!40}qCvM \\ 
  uniform.Gumbel & \cellcolor{red!40}BB & \cellcolor{red!20}ES & \cellcolor{blue!40}qAD & \cellcolor{blue!40}qCvM & \cellcolor{blue!40}FF & \cellcolor{blue!40}BR \\ 
  uniform.Galambos & \cellcolor{red!40}BB & \cellcolor{red!20}ES & \cellcolor{green!40}qAD & \cellcolor{blue!40}FF & \cellcolor{blue!40}qCvM & \cellcolor{blue!40}KSD \\
\hline
\end{tabular}
\caption{Power, continuous data, D=2, no estimation} 
\label{tblG.P.C2.nE}
\end{table}

%% file: tables/tblG.P.C2.E.tex
\begin{table}[H]
\centering
\begin{tabular}{lllllll}
  \hline
  \hline
normal-sigma.t & \cellcolor{red!40}KSD & ES & qK & BR & EP & RK \\ 
  Frank.Frank & \cellcolor{red!40}KSD & \cellcolor{red!20}RK & \cellcolor{green!40}EP & \cellcolor{green!40}BB & \cellcolor{green!40}ES & BR \\ 
  Clayton.Clayton & \cellcolor{red!40}qAD & \cellcolor{red!20}ES & \cellcolor{red!20}EP & \cellcolor{green!40}FF & \cellcolor{blue!40}BB & qKS \\ 
  Gumbel.Gumbel & \cellcolor{red!40}KSD & \cellcolor{green!40}EP & \cellcolor{green!40}ES & \cellcolor{blue!40}FF & BB & qCvM \\ 
  Galambos.Galambos & \cellcolor{red!40}qAD & \cellcolor{red!20}FF & \cellcolor{green!40}BB & \cellcolor{green!40}ES & \cellcolor{green!40}EP & qCvM \\ 
  HuslerReiss.HuslerReiss & \cellcolor{red!40}FF & \cellcolor{red!20}EP & \cellcolor{green!40}RK & BB & qCvM & qAD \\ 
  Joe.Joe & \cellcolor{red!40}KSD & \cellcolor{red!20}RK & \cellcolor{red!20}EP & \cellcolor{blue!40}FF & \cellcolor{blue!40}BB & qCvM \\ 
  mix.uniform.Frank & \cellcolor{red!40}BB & \cellcolor{red!20}ES & \cellcolor{red!20}RK & \cellcolor{green!40}FF & \cellcolor{blue!40}KSD & qAD \\
\hline
\end{tabular}
\caption{Power, continuous data, D=2, with estimation} 
\label{tblG.P.C2.E}
\end{table}

%% file: tables/tblG.P.C5.tex
\begin{table}[H]
\centering
\begin{tabular}{llllll}
  \hline
  \hline
uniform.uniform-diagonal-n & \cellcolor{red!40}qK & \cellcolor{green!40}BR & \cellcolor{blue!40}qAD & \cellcolor{blue!40}qKS & qK \\ 
  uniform.uniform-diagonal-b & \cellcolor{red!40}qK & \cellcolor{green!40}qAD & \cellcolor{blue!40}qKS & BR & qK \\ 
  normal.sigA & \cellcolor{red!40}BR & \cellcolor{blue!40}qKS & qK & qAD & BR \\ 
  normal.sigB & \cellcolor{red!40}BR & \cellcolor{blue!40}qKS & qK & qAD & BR \\ 
  t-ind.t-cor & \cellcolor{red!40}qAD & \cellcolor{green!40}BR & \cellcolor{blue!40}qKS & qK & qAD \\ 
  uniform.Frank & \cellcolor{red!40}qK & \cellcolor{red!20}BR & \cellcolor{red!20}qAD & \cellcolor{red!20}qKS & \cellcolor{red!20}qK \\ 
  uniform.Clayton & \cellcolor{red!40}qAD & \cellcolor{blue!40}qKS & qK & BR & qAD \\ 
  uniform.Gumbel & \cellcolor{red!40}qK & \cellcolor{green!40}qCvM & \cellcolor{blue!40}BR & \cellcolor{blue!40}qKS & qK \\
\hline
\end{tabular}
\caption{Power, continuous data, D=5, no estimation} 
\label{tblG.P.C5}
\end{table}

%% file: tables/tblG.P.H22.tex
\begin{table}[H]
\centering
\begin{tabular}{llllll}
  \hline
  \hline
uniform.uniform-diagonal-n & \cellcolor{red!40}qK & \cellcolor{green!40}BR & \cellcolor{blue!40}qAD & \cellcolor{blue!40}qKS & qK \\ 
  uniform.uniform-diagonal-b & \cellcolor{red!40}qK & \cellcolor{green!40}qAD & \cellcolor{blue!40}qKS & BR & qK \\ 
  normal.sigA & \cellcolor{red!40}BR & \cellcolor{blue!40}qKS & qK & qAD & BR \\ 
  normal.sigB & \cellcolor{red!40}BR & \cellcolor{blue!40}qKS & qK & qAD & BR \\ 
  t-ind.t-cor & \cellcolor{red!40}qAD & \cellcolor{green!40}BR & \cellcolor{blue!40}qKS & qK & qAD \\ 
  uniform.Frank & \cellcolor{red!40}qK & \cellcolor{red!20}BR & \cellcolor{red!20}qAD & \cellcolor{red!20}qKS & \cellcolor{red!20}qK \\ 
  uniform.Clayton & \cellcolor{red!40}qAD & \cellcolor{blue!40}qKS & qK & BR & qAD \\ 
  uniform.Gumbel & \cellcolor{red!40}qK & \cellcolor{green!40}qCvM & \cellcolor{blue!40}BR & \cellcolor{blue!40}qKS & qK \\
\hline
\end{tabular}
\caption{Power, continuous data, hybrid methods, nMC=2, D=2, no estimation} 
\label{tblG.P.H22}
\end{table}

%% file: tables/tblG.P.H52.tex
\begin{table}[H]
\centering
\begin{tabular}{llllll}
  \hline
  \hline
uniform.uniform-diagonal-n & \cellcolor{red!40}qK & \cellcolor{green!40}BR & \cellcolor{blue!40}qAD & \cellcolor{blue!40}qKS & qK \\ 
  uniform.uniform-diagonal-b & \cellcolor{red!40}qK & \cellcolor{green!40}qAD & \cellcolor{blue!40}qKS & BR & qK \\ 
  normal.sigA & \cellcolor{red!40}BR & \cellcolor{blue!40}qKS & qK & qAD & BR \\ 
  normal.sigB & \cellcolor{red!40}BR & \cellcolor{blue!40}qKS & qK & qAD & BR \\ 
  t-ind.t-cor & \cellcolor{red!40}qAD & \cellcolor{green!40}BR & \cellcolor{blue!40}qKS & qK & qAD \\ 
  uniform.Frank & \cellcolor{red!40}qK & \cellcolor{red!20}BR & \cellcolor{red!20}qAD & \cellcolor{red!20}qKS & \cellcolor{red!20}qK \\ 
  uniform.Clayton & \cellcolor{red!40}qAD & \cellcolor{blue!40}qKS & qK & BR & qAD \\ 
  uniform.Gumbel & \cellcolor{red!40}qK & \cellcolor{green!40}qCvM & \cellcolor{blue!40}BR & \cellcolor{blue!40}qKS & qK \\
\hline
\end{tabular}
\caption{Power, continuous data, hybrid methods, nMC=5, D=2, no estimation} 
\label{tblG.P.H52}
\end{table}

%% file: tables/tblG.P.H25.tex
\begin{table}[H]
\centering
\begin{tabular}{lllllll}
  \hline
  \hline
uniform.uniform-diagonal-n & \cellcolor{red!40}BG & \cellcolor{red!20}MMD & \cellcolor{green!40}CvM & \cellcolor{blue!40}AZ & \cellcolor{blue!40}NN5 & BF \\ 
  uniform.uniform-diagonal-b & \cellcolor{red!40}BG & \cellcolor{green!40}CvM & \cellcolor{blue!40}MMD & \cellcolor{blue!40}NN5 & AZ & FR \\ 
  normal-ind.normal-corA & \cellcolor{red!40}BG & \cellcolor{red!20}CvM & KS & MMD & FR & CF1 \\ 
  normal-ind.normal-corB & \cellcolor{red!40}K & \cellcolor{green!40}CvM & \cellcolor{green!40}NN5 & \cellcolor{blue!40}NN1 & \cellcolor{blue!40}MMD & KS \\ 
  t-ind.t-cor & \cellcolor{red!40}Ball & \cellcolor{green!40}CvM & MMD & NN5 & KS & AZ \\ 
  uniform.Frank & \cellcolor{red!40}BG & \cellcolor{green!40}AD & \cellcolor{green!40}CvM & \cellcolor{blue!40}AZ & \cellcolor{blue!40}BF & NN5 \\ 
  uniform.Clayton & \cellcolor{red!40}BG & \cellcolor{red!20}AZ & \cellcolor{red!20}MMD & \cellcolor{red!20}CvM & \cellcolor{red!20}NN5 & \cellcolor{red!20}BF \\ 
  uniform.Gumbel & \cellcolor{red!40}BG & \cellcolor{red!20}AD & \cellcolor{green!40}CvM & \cellcolor{blue!40}KS & \cellcolor{blue!40}AZ & \cellcolor{blue!40}FR \\
\hline
\end{tabular}
\caption{Power, continuous data, hybrid methods, nMC=1, D=5, no estimation} 
\label{tblG.P.H25}
\end{table}

%% file: tables/tblG.P.H55.tex
\begin{table}[H]
\centering
\begin{tabular}{lllllll}
  \hline
  \hline
uniform.uniform-diagonal-n & \cellcolor{red!40}Ball & \cellcolor{blue!40}AD & \cellcolor{blue!40}CvM & \cellcolor{blue!40}AZ & BF & FR \\ 
  uniform.uniform-diagonal-b & \cellcolor{red!40}BG & \cellcolor{red!20}AZ & \cellcolor{green!40}CvM & \cellcolor{green!40}AD & \cellcolor{blue!40}BF & NN5 \\ 
  normal-ind.normal-corA & \cellcolor{red!40}Ball & \cellcolor{red!20}CvM & \cellcolor{green!40}MMD & KS & NN5 & AZ \\ 
  normal-ind.normal-corB & \cellcolor{red!40}Ball & \cellcolor{red!20}MMD & \cellcolor{green!40}AD & \cellcolor{blue!40}AZ & \cellcolor{blue!40}KS & \cellcolor{blue!40}BF \\ 
  t-ind.t-cor & \cellcolor{red!40}Ball & \cellcolor{red!20}CvM & KS & MMD & NN5 & NN1 \\ 
  uniform.Frank & \cellcolor{red!40}BG & \cellcolor{green!40}AZ & \cellcolor{green!40}CvM & \cellcolor{blue!40}AD & \cellcolor{blue!40}BF & \cellcolor{blue!40}KS \\ 
  uniform.Clayton & \cellcolor{red!40}BG & \cellcolor{red!20}AD & \cellcolor{red!20}AZ & \cellcolor{red!20}BF & \cellcolor{red!20}MMD & \cellcolor{red!20}FR \\ 
  uniform.Gumbel & \cellcolor{red!40}BG & \cellcolor{green!40}CvM & \cellcolor{blue!40}KS & \cellcolor{blue!40}AZ & \cellcolor{blue!40}BF & AD \\
\hline
\end{tabular}
\caption{Power, continuous data, hybrid methods, nMC=5, D=5, no estimation} 
\label{tblG.P.H55}
\end{table}

%% file: tables/tblGH.P.2.tex
\begin{table}[H]
\centering
\begin{tabular}{lllllll}
  \hline
  \hline
uniform.uniform-diagonal-n & \cellcolor{red!40}BGT1 & \cellcolor{red!20}ESG & \cellcolor{red!20}MMDT5 & \cellcolor{red!20}MMDT1 & \cellcolor{red!20}EPT5 & \cellcolor{green!40}EST5 \\ 
  uniform.uniform-diagonal-b & \cellcolor{red!40}BGT5 & \cellcolor{red!20}MMDT5 & \cellcolor{red!20}ESG & \cellcolor{green!40}MMDT1 & \cellcolor{green!40}EST5 & \cellcolor{green!40}EPT5 \\ 
  normal-ind.normal-cor & \cellcolor{red!40}BGT5 & \cellcolor{red!20}ESG & \cellcolor{red!20}EPG & \cellcolor{red!20}EST5 & \cellcolor{green!40}qADG & \cellcolor{green!40}EPT1 \\ 
  t-ind.t-cor & \cellcolor{red!40}RKG & \cellcolor{green!40}EPT5 & \cellcolor{green!40}qADG & \cellcolor{green!40}ESG & \cellcolor{blue!40}qCvMG & \cellcolor{blue!40}EPT1 \\ 
  uniform.Frank & \cellcolor{red!40}BGT5 & \cellcolor{red!20}EPG & \cellcolor{red!20}ESG & \cellcolor{red!20}EST5 & \cellcolor{red!20}EPT5 & \cellcolor{green!40}MMDT1 \\ 
  uniform.Clayton & \cellcolor{red!40}BGT5 & \cellcolor{red!20}EPG & \cellcolor{red!20}MMDT5 & \cellcolor{red!20}qADG & \cellcolor{red!20}EPT5 & \cellcolor{red!20}EST5 \\ 
  uniform.Gumbel & \cellcolor{red!40}BGT5 & \cellcolor{red!20}MMDT5 & \cellcolor{red!20}ESG & \cellcolor{red!20}EST5 & \cellcolor{green!40}EPT5 & \cellcolor{green!40}MMDT1 \\ 
  uniform.Galambos & \cellcolor{red!40}BGT5 & \cellcolor{red!20}ESG & \cellcolor{red!20}MMDT5 & \cellcolor{red!20}EST5 & \cellcolor{red!20}EPT5 & \cellcolor{green!40}MMDT1 \\
\hline
\end{tabular}
\caption{Power, continuous data, gof+hybrid methods, D=2, no estimation} 
\label{tblGH.P.2}
\end{table}

%% file: tables/tblGH.P.5.tex
\begin{table}[H]
\centering
\begin{tabular}{lllllll}
  \hline
  \hline
uniform.uniform-diagonal-n & \cellcolor{red!40}BGT1 & \cellcolor{red!20}MMDT5 & \cellcolor{green!40}BRG & \cellcolor{green!40}ADT1 & \cellcolor{green!40}MMDT1 & \cellcolor{blue!40}qADG \\ 
  uniform.uniform-diagonal-b & \cellcolor{red!40}BGT5 & \cellcolor{red!20}MMDT5 & \cellcolor{green!40}ADT1 & \cellcolor{green!40}qADG & \cellcolor{green!40}AZT5 & \cellcolor{green!40}CvMT5 \\ 
  normal.sigA & \cellcolor{red!40}BGT1 & \cellcolor{green!40}ADT1 & \cellcolor{green!40}CvMT1 & \cellcolor{blue!40}qKSG & \cellcolor{blue!40}ADT5 & \cellcolor{blue!40}CvMT5 \\ 
  normal.sigB & \cellcolor{red!40}BallT5 & \cellcolor{blue!40}qKSG & \cellcolor{blue!40}CvMT5 & \cellcolor{blue!40}MMDT5 & ADT5 & qKG \\ 
  t-ind.t-cor & \cellcolor{red!40}BallT5 & \cellcolor{green!40}ADT5 & \cellcolor{green!40}BRG & \cellcolor{green!40}CvMT5 & \cellcolor{blue!40}qKSG & \cellcolor{blue!40}ADT1 \\ 
  uniform.Frank & \cellcolor{red!40}BGT5 & \cellcolor{red!20}BRG & \cellcolor{red!20}MMDT5 & \cellcolor{red!20}qADG & \cellcolor{red!20}qKSG & \cellcolor{red!20}qKG \\ 
  uniform.Clayton & \cellcolor{red!40}BGT5 & \cellcolor{red!20}AZT1 & \cellcolor{red!20}MMDT1 & \cellcolor{red!20}CvMT5 & \cellcolor{red!20}ADT5 & \cellcolor{red!20}AZT5 \\ 
  uniform.Gumbel & \cellcolor{red!40}BGT5 & \cellcolor{red!20}qADG & \cellcolor{green!40}qCvMG & \cellcolor{green!40}CvMT5 & \cellcolor{blue!40}KST5 & \cellcolor{blue!40}AZT5 \\
\hline
\end{tabular}
\caption{Power, continuous data, gof+hybrid methods, D=5, no estimation} 
\label{tblGH.P.5}
\end{table}

%% file: tables/tblG.P.D.nE.tex
\begin{table}[H]
\centering
\begin{tabular}{lllllllll}
  \hline
  \hline
uniform.uniform-diagonal-n & \cellcolor{red!40}K & \cellcolor{red!20}KL & \cellcolor{red!20}H & \cellcolor{red!20}TV & \cellcolor{blue!40}AD & CvM & KS & K \\ 
  uniform.uniform-diagonal-b & \cellcolor{red!40}K & \cellcolor{red!20}KL & \cellcolor{red!20}TV & \cellcolor{red!20}H & \cellcolor{blue!40}AD & CvM & KS & K \\ 
  normal-ind.normal-cor & \cellcolor{red!40}K & \cellcolor{red!20}H & \cellcolor{blue!40}TV & \cellcolor{blue!40}P & AD & CvM & KS & K \\ 
  t-ind.t-cor & \cellcolor{red!40}P & \cellcolor{red!20}H & \cellcolor{blue!40}TV & AD & KS & CvM & K & P \\ 
  uniform.Frank & \cellcolor{red!40}KS & \cellcolor{red!20}KL & \cellcolor{red!20}H & \cellcolor{red!20}TV & \cellcolor{blue!40}AD & \cellcolor{blue!40}CvM & K & KS \\ 
  uniform.Clayton & \cellcolor{red!40}K & \cellcolor{red!20}P & \cellcolor{red!20}H & \cellcolor{red!20}TV & \cellcolor{green!40}AD & \cellcolor{blue!40}CvM & KS & K \\ 
  uniform.Gumbel & \cellcolor{red!40}K & \cellcolor{red!20}KL & \cellcolor{red!20}H & \cellcolor{green!40}TV & \cellcolor{blue!40}AD & CvM & KS & K \\ 
  uniform.Galambos & \cellcolor{red!40}K & \cellcolor{red!20}P & \cellcolor{red!20}H & \cellcolor{red!20}TV & \cellcolor{blue!40}AD & \cellcolor{blue!40}CvM & KS & K \\
\hline
\end{tabular}
\caption{Power, discrete data, no estimation} 
\label{tblG.P.D.nE}
\end{table}

%% file: tables/tblG.P.D.E.tex
\begin{table}[H]
\centering
\begin{tabular}{lllllllll}
  \hline
  \hline
normal-sigma.t & \cellcolor{red!40}KS & \cellcolor{green!40}KL & AD & H & K & CvM & TV & KS \\ 
  Frank.Frank & \cellcolor{red!40}K & \cellcolor{red!20}KL & \cellcolor{red!20}H & \cellcolor{blue!40}TV & CvM & KS & AD & K \\ 
  Clayton.Clayton & \cellcolor{red!40}KS & \cellcolor{green!40}KL & H & TV & AD & K & CvM & KS \\ 
  Gumbel.Gumbel & \cellcolor{red!40}K & \cellcolor{red!20}KL & \cellcolor{green!40}H & \cellcolor{green!40}TV & AD & CvM & KS & K \\ 
  Galambos.Galambos & \cellcolor{red!40}K & \cellcolor{red!20}KL & \cellcolor{green!40}H & \cellcolor{blue!40}TV & CvM & AD & KS & K \\ 
  HuslerReiss.HuslerReiss & \cellcolor{red!40}K & \cellcolor{red!20}KL & \cellcolor{red!20}TV & \cellcolor{red!20}H & AD & CvM & KS & K \\ 
  Joe.Joe & \cellcolor{red!40}CvM & \cellcolor{red!20}KL & \cellcolor{red!20}H & \cellcolor{red!20}TV & AD & K & KS & CvM \\ 
  mix.uniform.Frank & \cellcolor{red!40}K & \cellcolor{red!20}KL & \cellcolor{red!20}TV & \cellcolor{red!20}H & AD & CvM & KS & K \\
\hline
\end{tabular}
\caption{Power, discrete data, with estimation} 
\label{tblG.P.D.E}
\end{table}

%% file: tables/tblT.P.C2.tex
\begin{table}[H]
\centering
\begin{tabular}{llllllllll}
  \hline
  \hline
NormalD2 & \cellcolor{red!40}BG & \cellcolor{red!20}EP & \cellcolor{red!20}MMD & \cellcolor{red!20}AZ & \cellcolor{green!40}BF & \cellcolor{blue!40}AD & \cellcolor{blue!40}NN5 & \cellcolor{blue!40}CvM & \cellcolor{blue!40}FR \\ 
  tD2 & \cellcolor{red!40}BG & \cellcolor{red!20}MMD & \cellcolor{red!20}ES & \cellcolor{green!40}AZ & \cellcolor{green!40}BF & \cellcolor{blue!40}AD & \cellcolor{blue!40}NN5 & \cellcolor{blue!40}CvM & \cellcolor{blue!40}K \\ 
  UniformMixtureD2 & \cellcolor{red!40}BG & \cellcolor{red!20}ES & \cellcolor{red!20}EP & \cellcolor{red!20}AZ & \cellcolor{blue!40}BF & \cellcolor{blue!40}NN5 & \cellcolor{blue!40}FR & \cellcolor{blue!40}CF1 & \cellcolor{blue!40}CF3 \\ 
  FrankD2 & \cellcolor{red!40}BG & \cellcolor{red!20}AZ & \cellcolor{green!40}ES & \cellcolor{green!40}EP & \cellcolor{green!40}BF & \cellcolor{blue!40}AD & Ball & CvM & KS \\ 
  ClaytonD2 & \cellcolor{red!40}BG & \cellcolor{red!20}BF & \cellcolor{red!20}MMD & \cellcolor{red!20}FR & \cellcolor{red!20}CF1 & \cellcolor{red!20}CF3 & \cellcolor{red!20}ES & \cellcolor{red!20}EP & \cellcolor{red!20}CF4 \\ 
  GumbelD2 & \cellcolor{red!40}BG & \cellcolor{red!20}AZ & \cellcolor{green!40}ES & \cellcolor{green!40}EP & \cellcolor{green!40}BF & \cellcolor{blue!40}AD & CvM & NN5 & FR \\ 
  GalambosD2 & \cellcolor{red!40}BG & \cellcolor{red!20}AZ & \cellcolor{red!20}ES & \cellcolor{red!20}EP & \cellcolor{green!40}BF & \cellcolor{blue!40}AD & \cellcolor{blue!40}NN5 & CvM & Ball \\ 
  HuslerReissD2 & \cellcolor{red!40}BG & \cellcolor{red!20}MMD & \cellcolor{red!20}ES & \cellcolor{red!20}EP & \cellcolor{red!20}AZ & \cellcolor{red!20}NN5 & \cellcolor{red!20}FR & \cellcolor{red!20}CF1 & \cellcolor{red!20}CF3 \\
\hline
\end{tabular}
\caption{Power, continuous data, dim=2} 
\label{tblT.P.C2}
\end{table}

%% file: tables/tblT.P.C5.tex
\begin{table}[H]
\centering
\begin{tabular}{lllllllll}
  \hline
  \hline
NormalD5 & \cellcolor{red!40}CF2 & \cellcolor{red!20}AZ & \cellcolor{red!20}K & \cellcolor{red!20}KS & \cellcolor{red!20}MMD & \cellcolor{green!40}CvM & \cellcolor{green!40}AD & \cellcolor{green!40}Ball \\ 
  tD5 & \cellcolor{red!40}BG & \cellcolor{red!20}AZ & \cellcolor{red!20}MMD & \cellcolor{green!40}Ball & \cellcolor{green!40}K & \cellcolor{blue!40}KS & \cellcolor{blue!40}AD & \cellcolor{blue!40}CvM \\ 
  FrankD5 & \cellcolor{red!40}BG & \cellcolor{red!20}AZ & \cellcolor{green!40}NN5 & \cellcolor{green!40}AD & \cellcolor{green!40}CvM & \cellcolor{green!40}BF & \cellcolor{blue!40}FR & \cellcolor{blue!40}CF1 \\ 
  ClaytonD5 & \cellcolor{red!40}BG & \cellcolor{red!20}MMD & \cellcolor{red!20}AZ & \cellcolor{red!20}K & \cellcolor{red!20}KS & \cellcolor{green!40}NN5 & \cellcolor{green!40}FR & \cellcolor{green!40}CF1 \\ 
  GumbelD5 & \cellcolor{red!40}BG & \cellcolor{red!20}MMD & \cellcolor{red!20}BF & \cellcolor{blue!40}K & \cellcolor{blue!40}NN5 & \cellcolor{blue!40}FR & \cellcolor{blue!40}CF1 & \cellcolor{blue!40}CF3 \\ 
  JoeD5 & \cellcolor{red!40}CvM & \cellcolor{red!20}NN5 & \cellcolor{red!20}AZ & \cellcolor{red!20}K & \cellcolor{red!20}FR & \cellcolor{red!20}CF1 & \cellcolor{red!20}CF3 & \cellcolor{green!40}NN1 \\ 
  UniformFrankD5 & \cellcolor{red!40}BG & \cellcolor{red!20}MMD & \cellcolor{red!20}CvM & \cellcolor{blue!40}NN5 & \cellcolor{blue!40}AZ & BF & KS & FR \\ 
  FrankClaytonD5 & \cellcolor{red!40}Ball & \cellcolor{red!20}CF2 & \cellcolor{red!20}CF4 & \cellcolor{blue!40}MMD & \cellcolor{blue!40}FR & \cellcolor{blue!40}CF1 & \cellcolor{blue!40}CF3 & \cellcolor{blue!40}AD \\
\hline
\end{tabular}
\caption{Power, continuous data, dim=5} 
\label{tblT.P.C5}
\end{table}

%% file: tables/tblT.P.D.tex
\begin{table}[H]
\centering
\begin{tabular}{lllllllll}
  \hline
  \hline
NormalD2 & \cellcolor{red!40}AZ & NN & AD & CvM & KS & K & BF & AZ \\ 
  tD2 & \cellcolor{red!40}BF & NN & K & KS & AD & CvM & AZ & BF \\ 
  UniformMixtureD2 & \cellcolor{red!40}BF & \cellcolor{green!40}NN & AD & K & CvM & KS & AZ & BF \\ 
  FrankD2 & \cellcolor{red!40}BF & \cellcolor{blue!40}NN & \cellcolor{blue!40}AD & CvM & KS & K & AZ & BF \\ 
  ClaytonD2 & \cellcolor{red!40}AZ & \cellcolor{red!20}NN & \cellcolor{red!20}K & \cellcolor{red!20}AD & \cellcolor{green!40}CvM & \cellcolor{green!40}KS & BF & AZ \\ 
  GumbelD2 & \cellcolor{red!40}BF & NN & AD & CvM & KS & K & AZ & BF \\ 
  GalambosD2 & \cellcolor{red!40}BF & AD & NN & CvM & K & KS & AZ & BF \\ 
  HuslerReissD2 & \cellcolor{red!40}BF & \cellcolor{red!20}NN & \cellcolor{green!40}AD & \cellcolor{green!40}K & \cellcolor{green!40}CvM & \cellcolor{blue!40}KS & AZ & BF \\
\hline
\end{tabular}
\caption{Power, discrete data} 
\label{tblT.P.D}
\end{table}